\documentclass[lettersize,journal]{IEEEtran}
\usepackage{amsmath,amsfonts}
\usepackage{algorithmic}
\usepackage{algorithm}
\usepackage{array}
\usepackage[caption=false,font=normalsize,labelfont=sf,textfont=sf]{subfig}
\usepackage{textcomp}
\usepackage{stfloats}
\usepackage{url}
\usepackage{verbatim}
\usepackage{xcolor}
\usepackage{graphicx}
\usepackage{cite}
\usepackage{booktabs} 

\usepackage{tcolorbox}
\tcbuselibrary{breakable}
\usepackage{float}

\usepackage{tikz}
\definecolor{CaptionColor}{RGB}{89, 91, 97}
\newcommand*\captionID[1]{\tikz[baseline=(char.base)]{
            \node[shape=rectangle,fill=white, text=CaptionColor, draw=CaptionColor,   
      line width=1pt,  inner sep= 1.2pt, minimum size=8pt,rounded corners=1pt] (char) {\textbf{#1}}}}

\newcommand{\techName}[1]{\textit{ReVis}}

\begin{document}

\title{\techName{}: Towards Reusable Image-Based Visualizations with MLLMs}


\author{
Xiaolin Wen, 
Changlin Li, 
Manusha Karunathilaka, 
Can Liu, 
Fangzhuo Jin, 
and Yong Wang
\thanks{X. Wen, C. Liu, and Y. Wang are with Nanyang Technological University. Email: xiaolin004@e.ntu.edu.sg, 
{can.liu,  yong-wang}@ntu.edu.sg. }
\thanks{C. Li is with Sichuan University. Email: same.lin@qq.com.}
\thanks{M. Karunathilaka is with Singapore Management University. Email: gmik.vidana.2023@phdcs.smu.edu.sg.}
\thanks{F. Jin is with Huazhong University of Science and Technology. Email: U202217231@hust.edu.cn.}
}

\markboth{Journal of \LaTeX\ Class Files,~Vol.~14, No.~8, August~2021}%
{Shell \MakeLowercase{\textit{et al.}}: A Sample Article Using IEEEtran.cls for IEEE Journals}


\maketitle

\begin{abstract}
Many expressive visualizations are shared online only as bitmap images, making them difficult to redesign or adapt to new data. Reusing such image-based visualizations requires substantial expertise and is often time-consuming, even for experienced visualization practitioners.
Existing work on reproducing visualizations often relies on structured SVG or specifications, supports limited visualization types, and offers limited flexibility for customization. 
To address these challenges, we present \techName{}, a human-AI collaboration approach that enables flexible reuse of image-based visualizations. 
First, a generic Domain-Specific language (DSL) is proposed to model complex visualizations and support both visualization decomposition and reproduction. 
Then, \techName{} employs an MLLM-based pipeline to parse an image-based visualization into the DSL, delineating its core visual structures and data-to-encoding mappings, and further reproduces the visualization from the DSL. 
Finally, \techName{} includes an interactive interface to allow users to upload visualization images, inspect reproduced results, update the underlying data, and customize visual encodings. 
A gallery of 40 visualizations demonstrates the expressiveness of the DSL, and a quantitative study evaluates the reproduction quality of \techName{} on these examples.
Two usage scenarios and user interviews with 16 visualization practitioners demonstrate the effectiveness of \techName{}.
\end{abstract}



\section{Introduction}
A vast number of visualization examples have been shared online through research publications, news articles, dashboards, and social media, yet most of them are only available as static images, which makes it difficult for users to directly reuse these designs for their own datasets, particularly when the designs are customized (e.g., composite visualizations~\cite{deng2023revisit}).
Although experienced visualization developers may employ tools such as Tableau or programming libraries like D3.js to reproduce a design, the process remains challenging and time-consuming, as it requires understanding and decomposing the visual structure, identifying the underlying visual encodings, and then manually re-implementing the visual design for new datasets.
For novice visualization practitioners, this challenge is even more pronounced, as they often lack the technical expertise or design experience necessary to reuse complex visualizations effectively~\cite {xie2025datawink}.

Prior studies have explored various strategies to streamline the visualization reuse process.
Many existing approaches rely on essential resources such as original SVG files or structured specifications.
For example, they deconstruct SVG-based visualizations, extract their data mappings, and convert them into reusable templates to support restyling and reuse~\cite{DeconstructingD3, harper2017converting, li2022structure, chen2023mystique, xie2025datawink}.
Bitmap-image-based approaches have also been explored to recover basic charts to enable redrawing or redesign~\cite{savva2011revision,poco2017reverse,jung2017chartsense,ying2024reviving}.
However, existing visualization reuse approaches still face three key challenges:
\textbf{C1. Heavy reliance on structured representations (e.g., SVG).} 
Many reuse methods assume access to structured resources such as SVG, enabling reliable extraction of chart components and mappings for reuse; however, such resources are often unavailable in practice~\cite{xie2025datawink,chen2023mystique}.
\textbf{C2. Limited coverage of visualization types.}
Existing reuse approaches often focus on basic chart types (e.g., bar, line, and scatter plots) but remain inadequate for customized visualizations that do not fit into a specific type~\cite{poco2017reverse,savva2011revision}.
\textbf{C3. Limited flexibility for customization. }
Existing reuse methods typically support data replacement and visual marks restyling, but provide little support for flexible customization, such as modifying the composition patterns or changing visual encodings beyond the extracted template structure~\cite{harper2017converting}.

To address these challenges, we first propose a \textbf{Domain-Specific Language (DSL)} designed for bridging bitmap-image-based visualization deconstruction, reproduction, and reuse.
(Throughout this paper, ``image-based'' refers to bitmap inputs (e.g., PNG/JPG) rather than vector graphics such as SVG.)
The DSL can capture the essential hierarchical visual
structure, data patterns, and visual encodings of visualizations, including those complex composite visualizations.
It enables flexible composition of visual marks while remaining expressive enough to support faithful regeneration and redesign.
Built upon the proposed DSL, we present \textbf{\techName{}}, a novel approach that leverages Multimodal Large Language Models (MLLMs) to enable image-based visualization reuse.
Given a visualization image, \techName{} automatically parses its underlying design into a structured DSL and reproduces the visualization via an MLLM-based pipeline.
To facilitate redesign and reuse in practice, \techName{} includes an interactive interface that enables users to inspect the reproduced visualizations, edit the DSL, and update the underlying data with ease.
For evaluation, we conduct a quantitative analysis of 20 basic charts and 20 composite visualizations that are directly generated by our MLLM-based pipeline, and further present a gallery of these examples to demonstrate the expressiveness of the DSL.
Two usage scenarios and in-depth user interviews with 16 visualization practitioners demonstrate that \techName{} effectively supports the reuse of image-based visualizations.
Our major contributions are as follows:

\begin{itemize}
    \item We introduce a domain-specific language (DSL) that enables the deconstruction and reproduction of visualizations, supports composite designs, and can be used independently for visualization generation and redesign.
    
    \item We propose a human-AI collaboration approach, \techName{},
    that employs an MLLM-based pipeline to reproduce the image-based visualizations and enables flexible redesign and reuse via an interactive interface. The interactive interface and example gallery have been published online: \textcolor{blue}{\textit{\url{https://revis-ui.github.io/ReVis/}}}. 

    \item We evaluate \techName{} via a quantitative evaluation on 40 visualizations, two usage scenarios, and user interviews with 16 visualization practitioners.

\end{itemize}



\section{Related Work}
Our work is related to \textit{Visualization Reuse}, \textit{Domain-Specific Language for Visualization}, and \textit{MLLM for Data Visualization}. 
In this work, we use the term \emph{LLMs} to broadly refer to both text-only LLMs and multimodal large language models (MLLMs), unless otherwise specified.


\subsection{Visualization Reuse}

Reusing existing designs or adapting them for new contexts is a common practice across many domains, including web design~\cite{kumar2011bricolage} and graphic design~\cite{warner2023interactive}; visualization design is no exception~\cite{Bako2023understanding,baigelenov2025visualization}.
Nevertheless, the reuse of visualizations is non-trivial in practice, as it often requires extracting the underlying data-to-encoding mappings and reconstructing them into structured representations, such as source code (e.g., D3.js~\cite{bostock2011d3}
) or declarative specifications (e.g., Vega-Lite~\cite{satyanarayan2016vega}), that enable flexible reuse and adaptation.

To facilitate this process, prior work has proposed automated methods for transforming SVG-based visualizations into reusable templates~\cite{DeconstructingD3, harper2017converting, li2022structure, chen2023mystique, xie2025datawink}. Leveraging SVG as input provides the advantage of a structured format, making it easier to extract design features. However, a key limitation is that in real-world scenarios, many visualizations are disseminated only as bitmap images rather than as SVG files. To overcome this, another line of research has investigated converting visualization bitmap images directly into code~\cite{savva2011revision, jung2017chartsense, poco2017reverse, ying2024reviving, li2025metal}. While promising, such image-based approaches have so far been restricted to 
only a few types of relatively simple visualizations,
including basic charts~\cite{li2025metal,yang2024chartmimic},graphs~\cite{song2022vividgraph,song2024gvvst}, and infographics~\cite{9585700,zhu2019towards},
limiting their applicability to more complex designs.

\techName{} takes the more available visualization bitmap images as input and enables the reuse of diverse and complex composite visualizations, thereby addressing the limitations of prior SVG-based and bitmap-image-based approaches.

\subsection{Domain-Specific Language for Visualization}


Domain-specific languages (DSLs)~\cite{fowler2010domain} for visualization provide tailored abstractions rather than low-level rendering or general-purpose programming constructs. 
Existing visualization DSLs span both declarative and procedural paradigms.

Declarative DSLs specify what visual representations should be produced, without prescribing how to construct them step-by-step.
General chart grammars such as Vega-Lite~\cite{satyanarayan2016vega} and Polaris~\cite{stolte2002polaris} represent visualizations as JSON-style specifications, supporting a broad space of generic charts.
Beyond generic charts, several DSLs target particular visualization types, for example, hierarchical data~\cite{li2020gotree}, unit visualization~\cite{park2017atom},
contour visualization~\cite{li2023contour}, density map~\cite{jo2018declarative}, scatterplot~\cite{tao2020kyrixs}, diagrams~\cite{pollock2024bluefish}, graph~\cite{hoffswell2018setcola},
and volume visualization~\cite{choi2014vivaldi}.
Other DSLs are procedural DSLs, and specify how to construct a visualization as a sequence of operations~\cite{liu2021atlas}.
For instance,
Atlas~\cite{liu2021atlas} is a grammar-based procedural language that generates visualizations.
Mascot~\cite{liu2024manipulable} is a grammar for procedural generation of data visualization by operations.


However, existing visualization DSLs typically assume fixed chart types and rely on predefined defaults, primarily focusing on chart specification or generation.
As a result, they offer limited support for constructing diverse and expressive composite visualizations, especially those requiring hierarchical organization of visualization content.

\subsection{MLLM for Data Visualization}

Recent advances in LLMs have enabled a broad range of visualization-related capabilities, including generating, understanding, and reusing visualizations.

Existing research~\cite{chen2024viseval, li2024visualization} has shown that LLMs can directly \textit{generate} charts from natural language when guided by grammars such as Vega-Lite~\cite{satyanarayan2016vega}.
Beyond generation, recent studies~\cite{hong2025llms, WangZWLW23}
have also examined LLMs as chart readers, focusing on visualization \textit{understanding}.
Charts-of-Thought~\cite{das2025charts} further introduces a prompting strategy that decomposes tasks to improve reasoning accuracy.
Beyond prompt engineering, SimVecVis~\cite{liu2025simvecvis} proposed a dataset and fine-tunes LLMs to better understand charts.
However, 
these approaches still cannot work for visualizations that are more expressive and complex.
Recent systems such as DataWink~\cite{xie2025datawink} and DataSway~\cite{xie2025datasway} extend this line of work by reverse-engineering SVG-based charts into reusable templates or animated representations through LLMs, enabling users to \textit{reuse} existing designs.
Nevertheless, DataWink is constrained to SVG-based visualizations and cannot handle complex visualizations with rich structural organization (e.g., composite visualizations~\cite{deng2023composite}).
Also, a large number of real-world visualizations are not available in vector formats; instead, they exist only as bitmap images and often contain composite structures.
Direct visualization generation or parsing techniques cannot support the reuse of such visualizations.

In this work, we leverage the multimodal parsing capabilities of LLMs 
and introduce a carefully designed multi-step pipeline that converts complex visualizations into a structured representation constrained by our DSL.

\section{Domain-Specific Language (DSL) for Visualization Reproduction}
This section presents the design goals and key components of our DSL, including the \textit{hierarchical container model} and \textit{data specification} of containers.

\subsection{Design Goals}
Existing visualization DSLs are typically expressed as declarative specifications (e.g., Vega-Lite~\cite{satyanarayan2016vega}). They often rely on predefined chart types and default settings, and their flat specifications cannot explicitly capture the hierarchical structure of visual elements. As a result, they have limited expressiveness for customized designs beyond supported chart templates and provide less flexibility for redesign.
Another line of visualization DSLs focuses on describing how visual elements are generated from data by specifying mappings from data attributes to visual channels (e.g., the Grammar of Graphics~\cite{wilkinson2011grammar}). However, these approaches generally depend on available data and construct visualizations by composing from individual marks to higher-level structures, which may not directly support deconstructing complete composite visualizations and converting them into reusable hierarchical representations.
Building on this survey of existing visualization DSLs~\cite{satyanarayan2016vega,li2020gotree, liu2021atlas, park2017atom,liu2024manipulable,pollock2024bluefish}, we derive the design goals for our DSL to better support automated visualization decomposition and reuse of customized visualizations:

\begin{itemize}
    \item \textbf{G1. Flexible Hierarchical Decomposition.} 
    Customized composite visualizations typically organize visual marks into a hierarchical structure~\cite{ying2024vaid}.
    To support image-based decomposition, the extracted representation should explicitly capture this hierarchy, from an overview of high-level components down to individual visual marks.
    Therefore, our DSL should provide flexible hierarchical expressiveness to represent visualizations at multiple levels of abstraction.
    Such hierarchy facilitates faithful reproduction and supports complex composite visualizations (e.g., juxtaposed, superimposed, and nested compositions) within a unified specification framework~\cite{deng2023revisit}.

    \item \textbf{G2. Abstract Data Specification.} 
    Recovering the original dataset from the visualization is unnecessary for reuse, but faithful reproduction relies on mocked data with similar patterns.
    In addition, users need a structured way to update the underlying data when adapting the visualization to new contexts.
    Thus, the DSL must describe both the abstract data structure and how visual elements are generated from the data, enabling faithful data generation for visualization reproduction and providing a clear format reference for adapting new datasets.
    
    \item \textbf{G3. Flexible Modification and Composition.} The DSL should enable flexible design modification by supporting visual components operating independently within its coordinate space, without being constrained by the global layout. This enables users to easily adjust spatial composition and arrangements. Moreover, the specification should not be limited to the visual elements present in the image but also accommodate data-driven variations, such as data-driven glyphs, ensuring that the reproduced visualization remains adaptable to different data contexts.

    \item \textbf{G4. Flexible Design Coverage.} 
    Visualizations for reuse can vary widely in structural composition and visual encoding, making it hard to rely on predefined chart types or default configurations.
    To support faithful regeneration across diverse designs, the DSL should explicitly specify how visual marks are composed and laid out, which visual channels encode data attributes, and what decorative styles are applied (e.g., stroke width and colors).
    Thus, the DSL must provide comprehensive structural, encoding, and styling specifications to enable direct visualization reproduction and subsequent adaptation.

\end{itemize}

\begin{figure*}[!t]
    \centering
    \includegraphics[width=0.85\textwidth]{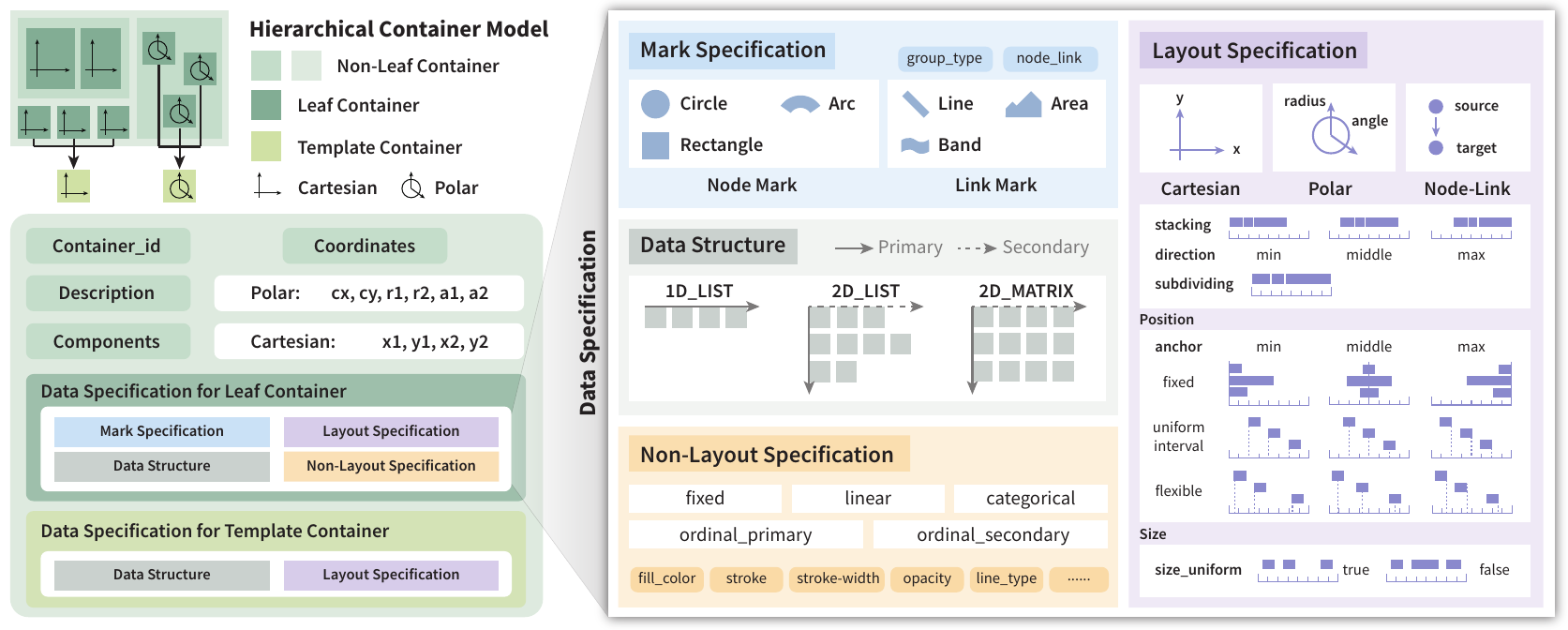}
    \caption{The domain-specific language (DSL) adopts a hierarchical container model, in which each container has its unique identifier and coordinates. Leaf containers include \textit{data specification} to mock data and render visual marks, encompassing \textit{mark specification}, \textit{data structure}, \textit{layout specification}, and \textit{non-layout specification}. Template containers contain only \textit{data structure} and \textit{layout specification} for generating repeated instances.
    }
    \label{fig: dsl}
\end{figure*}

Our DSL is designed to fulfill the above goals, as shown in Fig.~\ref{fig: dsl}. The details of our proposed DSL are described below.



\subsection{Hierarchical Container Model}
Following G1, our DSL employs a hierarchical container model to represent visualizations.
Specifically, it organizes visual elements into a hierarchy of containers, where each container corresponds to a composition of visual elements at a distinct level of abstraction.
Containers at the lowest level (\textit{\textbf{leaf containers}}) each contain a single type of visual mark governed by consistent data-mapping rules.
Containers at higher levels (\textit{\textbf{non-leaf containers}}) organize and compose containers, capturing common composition patterns such as nesting and superimposition~\cite{deng2023revisit}.
As shown in Fig.~\ref{fig: case2}\captionID{C}, the rectangles forming the bar chart and the reference line are first represented as two separate leaf containers (\textit{0-a-0} and \textit{0-a-1}). These leaf containers are then composed by a non-leaf container (\textit{0-a}), representing a unified visual component.
Each container has a unique identifier (\textit{container\_id}) and a coordinate system (Cartesian or Polar) specifying its occupying space and providing a base for independently organizing its inside sub-containers or visual marks (G3).
Leaf containers include a \textit{data specification} (elaborated in Sec.~\ref{sec:data_specification}) that details the visual marks and their data mappings, enabling data and code generation for visualization reproduction (G2).

As mentioned in G3, customized visualizations may include data-driven repetitive visual elements that share the same visual structure (e.g., glyphs~\cite{borgo2013glyph} and small multiples~\cite{tufte1990envisioning}), which can lead to redundant containers in the hierarchy of the DSL.
To simplify the DSL and automatically generate repetitive containers during reuse, we introduce a \textbf{\textit{template container}}.
Repetitive components are abstracted into a single template container, whose \textit{data specification} defines how its instance containers are instantiated and arranged based on the input data.
For example, all repetitive bar charts with a line in Fig.~\ref{fig: case2}\captionID{B} are represented by a template container (\textit{0-a}), whose internal containers (\textit{0-a-0} and \textit{0-a-1}) describe the visual encoding of one instance.
During reproduction, we instantiate all repeated containers according to the template container's \textit{data specification}, and render each instance using the \textit{data specification} of the leaf containers within that instance.
Users can easily reorganize these data-driven containers for reuse by modifying the template container's \textit{data specification}, rather than adjusting them one by one.


\subsection{Data Specification}\label{sec:data_specification}
To describe how data are generated for rendering visual marks and data-driven containers (G2), we introduce \textit{data specification}, specifying data patterns along with the detailed visual encoding without relying on chart type (G4).
For a leaf container, its \textit{data specification} consists of four components: (1) \textit{mark specification}, which describes the visual marks used in the container; (2) \textit{data structure}, which specifies how the data are organized; (3) \textit{layout specification}, which defines the spatial layout of visual marks; and (4) \textit{non-layout specification}, which specifies the encoding of decorative attributes such as color and stroke width.
Since the \textit{data specification} of a template container is only used to determine the spatial layout of its instance containers and is unrelated to mark rendering, it includes only \textit{data structure} and \textit{layout specification}.
Users can modify it to adjust the design and regenerate mocked data, allowing them to explore alternative designs without relying on real data (G3).
Details for each component are as follows:



\textbf{Mark Specification:}
Visual marks are building blocks of data visualization~\cite{liu2024manipulable}. 
Our DSL supports six types of visual marks (i.e., \textit{circle}, \textit{arc}, \textit{rectangle}, \textit{line}, \textit{band}, and \textit{area}), covering the major visual elements identified in prior research~\cite{liu2024manipulable,liu2021atlas,park2017atom}. 
Other marks can be derived from these fundamental types, such as a pie slice corresponding to an arc with a zero inner radius.
All the marks are categorized into two classes~\cite{munzner2014visualization}: \textit{node marks} and \textit{link marks}.
A \textit{node mark} is a geometric primitive whose visual properties (e.g., position, size, and shape) are primarily determined by a single data item (i.e., one data point with multiple attributes).
A \textit{link mark} encodes relationships among two or more data items, where its geometry is derived from tuples of entities (e.g., a sequence of control points defining a curve).
Accordingly, the mark specification defines the \textit{mark type} (e.g., circle and rectangle) as well as data-mapping attributes such as whether the link mark is determined by control points or other marks.

\textbf{Data Structure:} 
Our method requires inferring the data structure to generate data for rendering visual marks and to determine the coordinates of template container instances.
In this context, each node mark or template container instance is associated with a single data item, whereas each link mark corresponds to multiple data items that define its control points.
\textit{Data structure} specifies how these data items are organized, which we infer from the spatial layout of the corresponding visual elements (i.e., node marks, link-mark control points, and template container instances).
Considering the grouping and regularity patterns of visual elements in 2D space (e.g., rectangles in a matrix or in a stacked bar chart), we define three data types: \textit{1D\_list}, \textit{2D\_matrix}, or \textit{2D\_list}, as shown in Figure~\ref{fig: dsl}. 
\textit{1D\_list} represents a collection of data items without visual grouping patterns, where each data item corresponds to an independent visual element, such as the circles in a scatterplot. A \textit{2D\_matrix} describes data organized into groups of equal size, where each group forms a complete visual entity. Typical examples include a line chart, in which each line is defined by a fixed number of control points, or a stacked bar chart, in which each group contains the same number of stacked bars. Different from \textit{2D\_matrix}, a \textit{2D\_list} is used when the data items are grouped, but different groups have varying sizes.
For instance, a dot plot contains multiple categories of dots, but each category may contain a different number of dots.
We use \textit{primary} and \textit{secondary} dimensions to further define how the layout along axes corresponds to the data structure. 
The \textit{primary} dimension specifies how different data point groups
are distributed, corresponding to the outer grouping structure (e.g., categories along the x-axis in stacked charts). 
The \textit{secondary} dimension specifies how individual data items within each group are arranged, determining the internal composition of each group (e.g., bars within a single stacked bar).
The \textit{primary} and \textit{secondary} dimensions may coincide. For example, in a grouped bar chart, both groups and bars within each group are positioned along the x-axis. The \textit{primary} dimension can also span two axes, as in scatter plots, where marks are positioned jointly along the x- and y-axes.
The \textit{number} attribute of \textit{primary} dimension or \textit{secondary} dimension indicates the number of elements (i.e., groups or items within groups) along each dimension, facilitating accurate data generation and layout reasoning. 


\textbf{Layout Specification:} 
Visual marks and containers can have shape-specific geometric parameters, making it difficult to describe them using a unified representation (e.g., rectangles use $(x, y, width, height)$ while circles use $(c_x, c_y, r)$).
To address this challenge, our DSL introduces a unified \textit{layout\_specification} that standardizes layout representations across node marks, link-mark control points, and containers, enabling a consistent and data-driven description of spatial organization.
It abstracts individual geometric parameters into two visual channels: \textbf{\textit{position}} and \textbf{\textit{size}}.
To support diverse layouts in visualizations, the DSL specifies layouts independently along each dimension and then combines them into a full 2D layout.
In each dimension, we first check whether visual elements are stacked, because under stacking, \textit{position} and \textit{size} become interdependent and must be jointly determined to ensure non-overlapping placement.
When \textit{stacking} is set to \textit{true}, elements are stacked along the current dimension; we then use \textit{stacking\_direction} to specify the stacking order and \textit{subdividing} to indicate whether the stacked elements subdivide the entire available range.
In contrast, when $stacking$ is \textit{false}, \textit{position} and \textit{size} are specified independently. 
For \textit{position}, the \textit{anchor} attribute specifies the reference point for positioning (\textit{min}, \textit{max}, or \textit{middle}).
The \textit{anchor\_distribute} attribute controls how anchors are positioned along an axis, which has three options: \textit{fixed} (same position), \textit{uniform\_interval} (positioned equally), and \textit{flexible} (decided by data). 
When \textit{uniform\_interval} is used, \textit{anchor\_start} and \textit{anchor\_interval} specify the starting position and spacing between consecutive elements. 
For \textit{size}, the \textit{size\_uniform} and \textit{size\_range} attributes specify whether elements share the same size and define the proportional range of element sizes, expressed as relative percentages of the full axis length.

\textbf{Non-layout Specifications:} The \textit{non\_layout\_specification} defines attributes irrelevant to position and size, such as \textit{fill}, \textit{stroke}, \textit{stroke\_width}, and \textit{opacity}. 
Each attribute follows a unified schema with a \textit{scale} attribute that determines how values are assigned: \textit{fix}, \textit{linear}, \textit{ordinal\_primary}, \textit{ordinal\_secondary}, or \textit{categorical}. \textit{Fixed} uses a constant value; \textit{linear} interpolates within a numeric or color range;
\textit{ordinal\_primary} and \textit{ordinal\_secondary} mean that the value is encoded by group index or item index within each group; and \textit{categorical} indicates that the value is randomly selected from a set of options. 


\begin{figure*}[!t]
    \centering
    \includegraphics[width=0.9\textwidth]{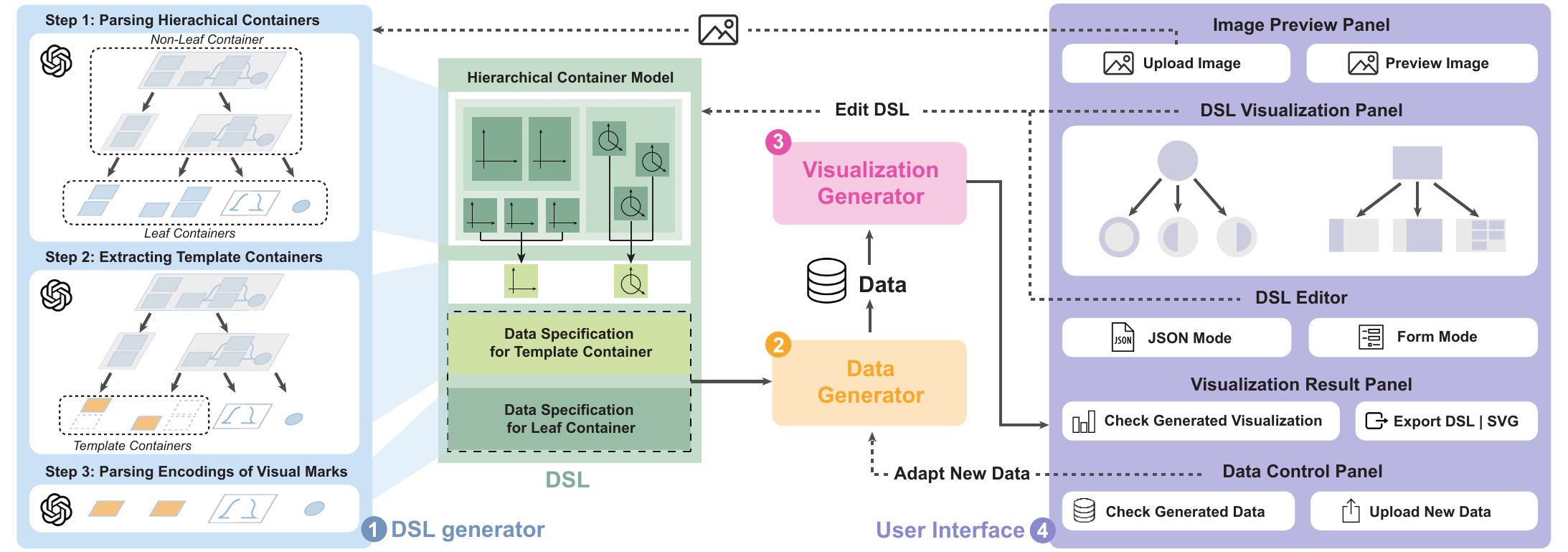}
    \caption{The MLLM-based pipeline for visualization reproduction first parses an input visualization image into a structured DSL representation using a DSL generator. It then generates data through a Data Generator and renders the visualization via a Visualization Generator. Finally, the \techName{} interface presents the generated results and supports DSL editing as well as customized data adaptation.
    }
    \label{fig:pipeline}
\end{figure*}

\begin{figure*}[!t]
    \centering
    \includegraphics[width=0.96\textwidth]{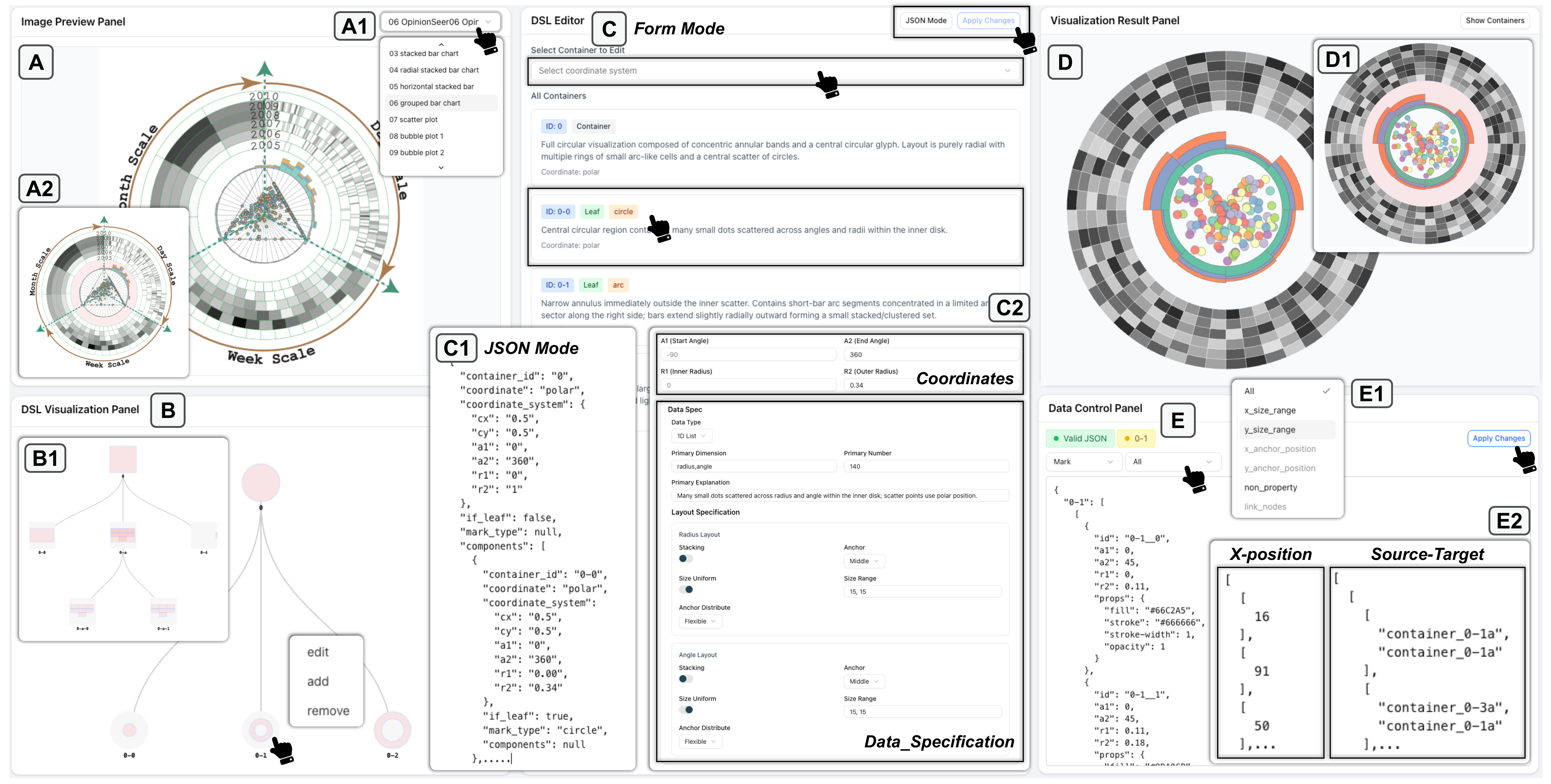}
    \caption{The \techName{} interface consists of five panels: (A) the Image Preview Panel, which allows users to upload and preview visualization images; (B) the DSL Visualization Panel, which provides an overview of how the DSL decomposes the visual design into hierarchical containers; (C) the DSL Editor, which enables users to edit the DSL to modify the visual design; (D) the Visualization Result Panel, which displays the visualization generated from the DSL; and (E) the Data Control Panel, which allows users to apply their own data to the visualization.
    }
    \label{fig: interface}
\end{figure*}

\section{\techName{}}\label{sec:pipeline}
\techName{} employs an MLLM-based pipeline for visualization reproduction (Fig.~\ref{fig:pipeline}).
It consists of four components:
\textit{DSL Generator} generates the DSL based on the visualization image,
\textit{Data Generator} generates the data based on the DSL,
\textit{Visualization Generator} reproduces the visualization using the DSL and the generated data,
 and an interactive \textit{User Interface} supports users to flexibly edit and reuse the design.



\subsection{DSL Generator} \label{sec: dsl generator}
The DSL generator parses a visualization image into the proposed DSL using an MLLM (ChatGPT-5\footnote{\url{https://platform.openai.com/}}) that has shown strong visual reasoning abilities. There are three main steps:

\textbf{Step 1: Parsing Hierarchical Containers.}
To decompose the visualization into a hierarchical container model that represents the compositional structure, we provide the MLLM with the input image and a crafted prompt that guides it to parse the image step by step based solely on the visual composition of visual marks, while ignoring decorative elements (e.g., axes and legends) and the underlying data semantics.
The prompt instructs the model to first identify the coordinate system (Cartesian or polar), and then recursively partition the visualization into sub-containers according to visual cues in the image, such as juxtaposition, grid, concentric, or angular sections. 
Each sub-container is assigned a distinct coordinate subspace and further decomposed, if possible, into finer-grained components until reaching indivisible leaf containers, i.e., a type of atomic visual marks (e.g., rectangles, arcs, circles, lines).
This hierarchical containerization not only preserves the structural layout of the visualization but also provides a robust intermediate representation for subsequent editing and regeneration.
Step 1 collects the basic structure of DSL with the containers' coordinates and textual descriptions.


\textbf{Step 2: Extracting Template Containers.}
We then provide the MLLM with the Step-1 output, the image, and a follow-up prompt to identify and merge repetitive containers into a template container, reducing redundancy and enabling data-driven container extraction. 
Specifically, the prompt instructs the MLLM to traverse the container hierarchy in a top-down manner, detect containers with similar coordinate shapes and descriptions.
All repeated containers in the DSL are replaced by a template container, whose coordinates correspond to the outer boundary enclosing the merged containers in the parent coordinate system.
Following our DSL definitions in the prompt, the MLLM is instructed to generate a \textit{data\_specification} for the template container that captures the data patterns and layouts of the merged containers, facilitating subsequent reproduction.
Step 2 collects the simplified DSL with the template data specifications.


\textbf{Step 3: Parsing Encodings of Visual Marks.}
Finally, we traverse all leaf containers in the DSL. For each leaf container,
we provide
an individual prompt to guide the MLLM in generating the corresponding \textit{data\_specification}. 
Before employing such a design, we also experimented with
using the same prompt for the MLLM to process all leaf containers, 
but it resulted in a lower accuracy and occasionally led to timeouts.
The prompt for each leaf container includes the image, the output from Step~2, the target \textit{container\_id}, and instructions based on our DSL definitions.
The MLLM is instructed to first locate the target visual component using its \textit{container\_id} and description, and then determine the corresponding \textit{mark\_specification} and \textit{data\_structure}. Next, it reasons step by step to infer the \textit{layout\_specification} and \textit{non\_layout\_specification} attributes, following the required output format.
By combining the \textit{data\_specification} of all leaf containers, Step 3 collects the final DSL for visualization reproduction, which serves as the foundation for generating data and faithfully reconstructing the visualization.

\subsection{Data Generator}

To reproduce the visual design of a given visualization image for editing and reuse, it is not necessary to extract the exact underlying data from the visualization image.
Instead, we identify data patterns that are specified by \textit{data\_specification} in the DSL (extracted from Steps 2 and 3 in Sec.~\ref{sec: dsl generator})
and then generate appropriate mocked data accordingly.
We implement a Data Generator to parse the \textit{data\_specification} of containers and further generate data accordingly using rule-based procedures.
Specifically, the \textit{data specification} of a template container is used to generate the coordinate data for its instance containers,
while each leaf container's specification guides the generation of a one- or two-dimensional list of objects, with each object corresponding to a single visual mark.
The attributes of these objects encode the properties required to render the marks as SVG elements, such as width, height, position, and color of a rectangle. 
When users modify data specifications, the Data Generator automatically regenerates and updates the corresponding data for visualization.

\subsection{Visualization Generator}
The Visualization Generator is designed to reproduce the bitmap-image-based visualization using the DSL and the mocked data generated by the Data Generator. It first maps the coordinate system of each container onto the canvas and constructs scaling functions to map data values into the coordinates. 
Subsequently, it invokes the corresponding drawing functions
to render the composite visualization by considering the specified mark types,
where the drawing functions are implemented using D3.js.

\begin{figure*}[!t]
    \centering
    \includegraphics[width=0.9\textwidth]{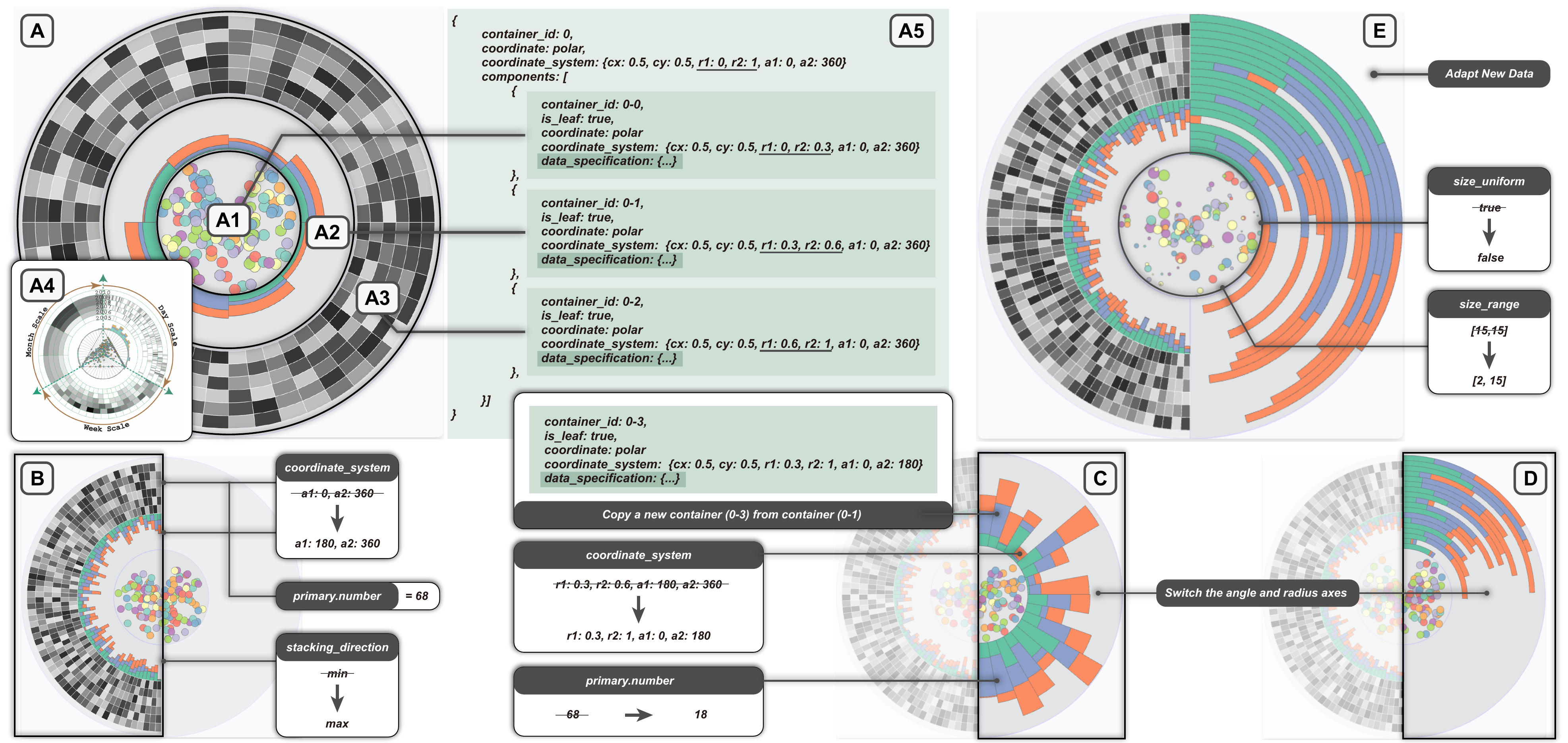}
    \caption{With \techName{}, a visualization designer can redesign a visualization from a static image. She first uploads a reference image (A4) to obtain the extracted DSL and the regenerated visualization (A). By interactively editing the DSL in the \textit{DSL Editor} (B-E), she explores different design variations, creates a customized design, and finally applies her own data to it.
    }
    \label{fig: case1}
\end{figure*}

\subsection{\techName{} Interface} \label{sec: interface}
We developed an interactive interface\footnote{\url{https://revis-ui.github.io/ReVis/#/editor}} that enables users to check and reuse the reproduced visualizations smoothly, as shown in Fig.~\ref{fig: interface}, which consists of five panels:

\textbf{Image Panel} (Fig.~\ref{fig: interface}\captionID{A}) serves as the entry point of \techName{}, allowing users to upload the visualization image they intend to reuse (Fig.~\ref{fig: interface}\captionID{A1}). 
Once uploaded, the target image-based visualization is displayed in the panel, and \techName{} automatically executes the visualization reproduction process described in Sec.~\ref{sec:pipeline}, presenting the extracted DSL (Fig.~\ref{fig: interface}\captionID{C}) and the reproduced visualization (Fig.~\ref{fig: interface}\captionID{D}).

\textbf{Visualization Result Panel} (Fig.~\ref{fig: interface}\captionID{D}) displays the visualization generated from the DSL. Whenever the DSL or underlying data is modified, the panel automatically updates to reflect the changes, allowing users to immediately verify the impact of their edits on the visualization design.

\textbf{DSL Editor} (Fig.~\ref{fig: interface}\captionID{C}) displays the generated DSL in JSON format (Fig.~\ref{fig: interface}\captionID{C1}), allowing users to refine the design by editing the DSL directly. 
To avoid overwhelming users with the full JSON structure, we also provide a Form Mode (Fig.~\ref{fig: interface}\captionID{C2}), where key parts, like coordinates and \textit{data\_specification}, of the DSL can be modified through intuitive interactions such as drop-down menus, input boxes, and buttons, making the editing process more accessible and user-friendly.
By clicking certain containers and editing their DSL, users can flexibly change the composition, visual encodings, and data patterns of visual components and directly preview the expected design in the Visualization Result Panel.


\textbf{DSL Visualization Panel} (Fig.~\ref{fig: interface}\captionID{B}) helps users understand how \techName{} decomposes a visualization into hierarchical containers and maps these containers to the original image. Specifically, the panel presents a tree visualization that depicts the container hierarchy, where each node corresponds to a container. Rectangular nodes indicate Cartesian coordinate systems (Fig.~\ref{fig: interface}\captionID{B1}), while circular nodes represent polar coordinate systems (Fig.~\ref{fig: interface}\captionID{B}).
Within each node, a shadowed shape illustrates the container’s relative position and area, enabling users to intuitively locate the corresponding visual components. To further strengthen the correspondence among the input image, the DSL, and the generated visualization, the panel supports interactive linking: hovering over a container highlights the corresponding region simultaneously in the original image (Fig.~\ref{fig: interface}\captionID{A2}) and the reproduced visualization (Fig.~\ref{fig: interface}\captionID{D1}).
Users can also remove containers or add new sub-containers via right-click interactions, and click on a container to inspect its associated DSL in the \textit{DSL Editor}.

\textbf{Data Control Panel} (Fig.~\ref{fig: interface}\captionID{E}) enables users to interactively apply their own data to the generated visualizations. After selecting a container, the panel displays the mocked data derived from its \textit{data\_specification}. Users can replace this mocked data with their own data by following the same data format and immediately apply the changes.
In addition, users can selectively update individual attributes (Fig.~\ref{fig: interface}\captionID{E1}) to partially replace the data, such as modifying x-position values or specifying source–target pairs for link marks (Fig.~\ref{fig: interface}\captionID{E2}).

\section{Usage Scenarios}
\label{sec: case}
This section presents two usage scenarios to demonstrate how \techName{} supports regenerating visualizations from images, refining visual designs, and adapting them to new data.

\subsection{Usage Scenario 1: Re-designing existing visualizations}
Alice is a visualization designer who comes across a compelling visualization in a paper~\cite{wu2010opinionseer} (Fig.~3\captionID{A4}). Inspired by its effective visual design, she wants to adapt it for her own dataset with similar analytical goals. However, the visualization is only available as a static image, making it costly to recreate from scratch. She therefore turns to \techName{} to explore and adapt the design.

\textbf{Reusing the original design.}
Alice uploads a screenshot of the visualization from the paper to \techName{}. After about ten minutes, \techName{} returns the extracted DSL representation of the design (Fig.~\ref{fig: case1}\captionID{A5}) and renders the corresponding visualization result (Fig.~\ref{fig: case1}\captionID{A}).
The design is parsed into three concentric annular bands: an inner scatter plot (Fig.~\ref{fig: case1}\captionID{A1}), a middle stacked radial bar chart (Fig.~\ref{fig: case1}\captionID{A2}), and an outer gridded heatmap (Fig.~\ref{fig: case1}\captionID{A3}), corresponding to containers \textit{0-0}, \textit{0-1}, and \textit{0-2}, respectively.
She decides to use only the left half of the stacked bars and the heatmap, and thus modifies the coordinate systems of containers \textit{0-1} and \textit{0-2} from ``\textit{a1: 0 – a2: 360}'' to ``\textit{a1: 180 – a2: 360}.''
Since Alice wants to show different attributes of the same objects along the angular axis, she adjusts the \textit{primary.number} of both containers to align the object counts, and changes the stacking direction of container \textit{0-1} so that the two containers stack together more compactly.
After applying these adjustments in the \textit{DSL Editor}, \techName{} renders the updated design, as shown in Fig.~\ref{fig: case1}\captionID{B}.

\textbf{Customizing her new design.}
Alice duplicates container \textit{0-1} as a new container \textit{0-3}, adjusts its coordinate to occupy the available space on the right half, and updates the \textit{primary.number} to match her data, as shown in Fig.~\ref{fig: case1}\captionID{C}.
Since Alice wants to compare the lengths of different stacked bars, the current angular layout makes such comparisons difficult due to misalignment. She therefore reorients the stacked bars along the radial axis to enable easier visual comparison. By simply exchanging the \textit{layout specifications} of ``radius'' and ``angle'' in the \textit{DSL Editor}, she obtains a satisfactory design (Fig.~\ref{fig: case1}\captionID{D}).
Then, she modifies the \textit{size\_uniform} and \textit{size\_range} of container \textit{0-0}, transforming the scatter plot into a bubble chart to encode an additional data dimension.
Finally, by uploading her data for each container following the reference format in the \textit{Data Control Panel}, she successfully builds a customized visual design (Fig.~\ref{fig: case1}\captionID{E}) based on an image-based visualization and applies her own data using \techName{}.

\begin{figure*}[!t]
    \centering
    \includegraphics[width=0.96\textwidth]{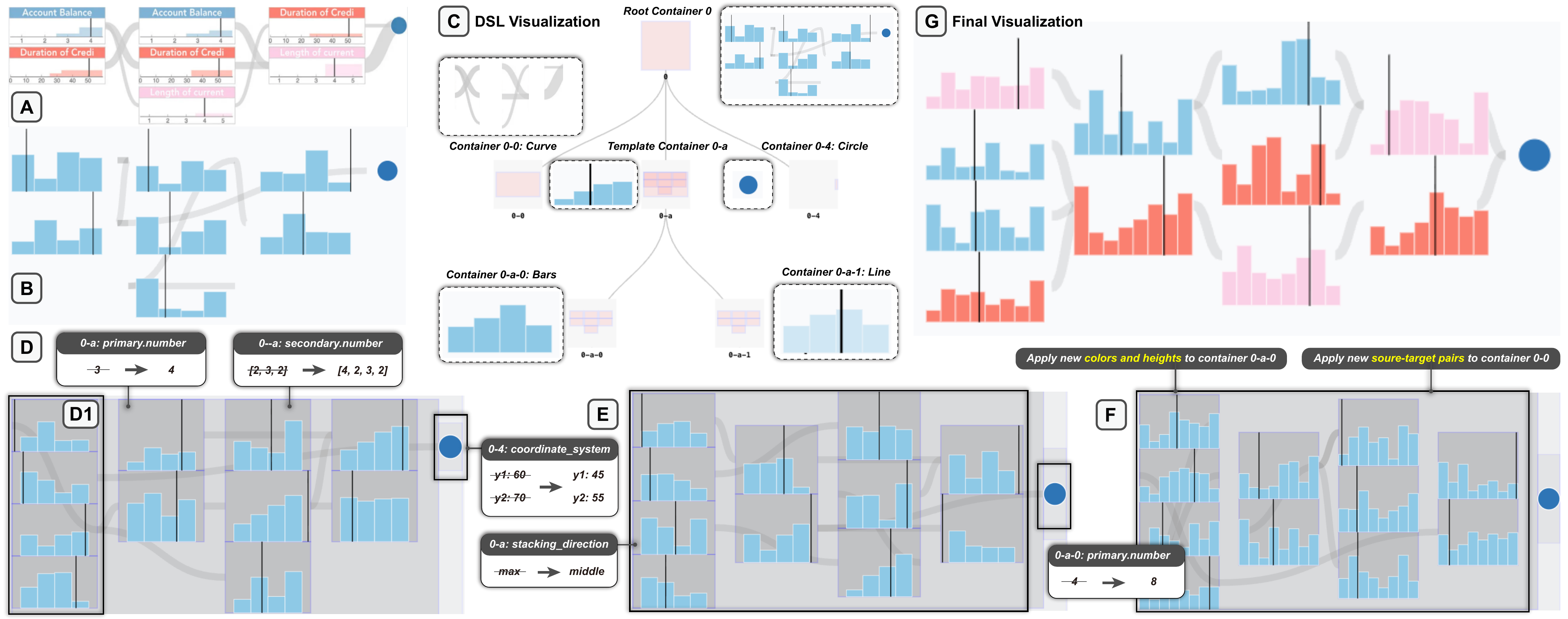}
    \caption{With \techName{}, a data analyst can flexibly restructure an existing image-based visualization (A) and apply his own data to the redesigned result. By editing the \textit{data\_specification} of the template container, he modifies the structure and layout of the repetitive bar–line components (D-F), and by following the exemplar data format, he interactively customizes bar colors and curve linking patterns (G).
    }
    \label{fig: case2}
\end{figure*}

\subsection{Usage Scenario 2: Reusing data-driven templates}
Mike is a data analyst who wants to reuse a composite visualization design from a paper~\cite{zhao2018iforest} for his own project, as shown in Fig.~\ref{fig: case2}\captionID{A}. 
However, the linked repetitive bar charts are difficult for him to implement, as he has limited experience with D3~\cite{bostock2011d3} and has only created simple charts. 
Mike wants to use \techName{} to reuse this image-based visual design.

\textbf{Adjusting the data-driven template containers.} 
After uploading the image of the target design to \techName{}, Mike obtains the regenerated visualization shown in Fig.~\ref{fig: case2}\captionID{B}. The \textit{DSL Visualization Panel} clearly reveals the hierarchical structure of the design (Fig.~\ref{fig: case2}\captionID{C}): the root container represents the overall layout; containers \textit{0-0} and \textit{0-4} correspond to the curves and the final circle; the template container \textit{0-a} captures the repetitive, data-driven bar–line components; and its sub-containers \textit{0-a-0} and \textit{0-a-1} represent the bars and the line, respectively.
By editing the \textit{data\_structure} of the template container \textit{0-a} in the \textit{DSL Editor}, Mike adds a new column of four instances (i.e., small bar charts with a line), as shown in Fig.~\ref{fig: case2}\captionID{D1}.
He then restructures the design into a middle-stacked layout (Fig.~\ref{fig: case2}\captionID{E}) by setting the \textit{stacking\_direction} of the template instances to \textit{middle} and repositioning the final circle through the \textit{coordinate\_system} of container \textit{0-4}.

\textbf{Applying his own data.}
To adapt the design to his data, Mike increases the number of bars from 4 to 8 by modifying the \textit{data\_structure} of container \textit{0-a-0}, as shown in Fig.~\ref{fig: case2}\captionID{F}. He then notices that the bar colors are fixed and the links between containers are randomly assigned. Using the \textit{Data Control Panel}, which provides the standard data format, Mike replaces the placeholder data with his own, specifying bar colors and source–target pairs for the curves. The resulting visualization is shown in Fig.~\ref{fig: case2}\captionID{E}.
With \techName{}, Mike can flexibly restructure the original design by reusing data-driven template containers and customize linking and visual encoding by replacing the example data through the \textit{Data Control Panel}.



\section{Quantitative Evaluation and Gallery} 
We conducted a quantitative study of reproduced visualizations by using \techName{} to demonstrate its effectiveness. 
We first constructed a gallery of bitmap-based visualization images and reproduced them using our pipeline without manual intervention, and then assessed the accuracy of the reproduced results.





\subsection{Gallery}\label{sec: gallery}
To evaluate the expressiveness of \techName{}, we constructed a visualization gallery\footnote{\url{https://revis-ui.github.io/ReVis/#/}} of 20 basic charts and 20 composite visualizations. 
The original visualization images are collected from the Vega-Lite Gallery\footnote{\url{https://vega.github.io/vega-lite/examples/}}
 and prior design studies~\cite{wu2010opinionseer,zhao2018iforest,yue2018bitextract,chen2016dropoutseer,xu2019clouddet,gratzl2013lineup,shen2016nameclarifier,xu2018ensemblelens}.
 For each original visualization image, we further generated the corresponding DSL specifications and a reproduced visualization by using \techName{}.
 To assess our DSL generation approach, we conducted a quantitative evaluation on both basic charts (Sec.~\ref{sec: basic_evaluation}) and composite visualizations (Sec.\ref{sec: composite_evaluation}).

\subsection{Basic charts}\label{sec: basic_evaluation}
The basic charts in our gallery cover diverse chart types, including bar charts, scatter plots, bubble plots, histograms, pie and donut charts, line and area charts, as well as more specialized visualizations such as parallel coordinates, radar charts, and mosaic plots.
For each visualization, we manually constructed a ground truth DSL that precisely describes the chart's visual encoding, data structure, and layout properties, serving as the reference standard against which we compare our automatically generated outputs.
Since visualization specifications are conditional, only attributes applicable to a given chart are evaluated. For example, stacking direction is assessed only when stacking is enabled, and polar-specific properties are evaluated only for radial charts. This results in a varying number of evaluated attributes across charts but ensures a fair comparison.
For each chart, we compute the accuracy as the proportion of correctly generated attributes among all applicable attributes, where an attribute is considered correct if it exactly matches the ground truth. 


\textbf{Results:}
We report both per-chart accuracy and the overall accuracy across 20 basic charts in Table~\ref{tab:evaluation_results}.
Overall, our approach achieved an accuracy of 94.8\%, correctly predicting 331 out of 349 applicable attributes. 
Most cases fully matched the ground-truth specifications, while the remaining 7 cases exhibited mismatches in color encoding (\#04), mark size (\#18), or layout attributes (\#05, \#09, \#12, \#13).
In the radial stacked bar chart, our method incorrectly identified the ordinal color scheme within each stacked bar as randomly assigned categorical colors. We attribute this error to the varying bar sizes in the radial layout, which make it difficult for the MLLM to recognize the underlying ordinal color encoding.
For layout attributes, our method may misidentify a \textit{uniform\_interval} anchor distribution as \textit{flexible} when alignment patterns are unclear in two-dimensional layouts (\#09, \#10, \#12) or when the data contain missing values (\#10, \#13).
Importantly, these issues can be easily addressed by modifying the corresponding attributes in the DSL through interactions supported by \techName{}.

\begin{table}[htbp]
\centering
\caption{Evaluation Results of Basic Charts}
\label{tab:evaluation_results}
\footnotesize
\begin{tabular}{@{}clcccc@{}}
\toprule
\textbf{\#} & \textbf{Chart Type} & \textbf{Acc. (\%)} & \textbf{Match} & \textbf{Mismatch} & \textbf{Total} \\
\midrule
01 & Simple Bar Chart & 100.0 & 18 & 0 & 18 \\
02 & Radial Bar Chart & 100.0 & 17 & 0 & 17 \\
03 & Stacked Bar Chart & 100.0 & 18 & 0 & 18 \\
04 & Radial Stacked Bar Chart & 94.4 & 17 & 1 & 18 \\
05 & Horizontal Stacked Bar & 83.3 & 15 & 3 & 18 \\
06 & Grouped Bar Chart & 100.0 & 18 & 0 & 18 \\
07 & Scatter Plot & 100.0 & 18 & 0 & 18 \\
08 & Bubble Plot 1 & 100.0 & 17 & 0 & 17 \\
09 & Bubble Plot 2 & 72.2 & 13 & 5 & 18 \\
10 & 2D Histogram Scatterplot & 84.2 & 16 & 3 & 19 \\
11 & 2D Histogram Heatmap & 100.0 & 19 & 0 & 19 \\
12 & Strip Plot & 77.8 & 14 & 4 & 18 \\
13 & Dot Plot & 94.7 & 18 & 1 & 19 \\
14 & Pie Chart & 100.0 & 17 & 0 & 17 \\
15 & Donut Chart & 100.0 & 17 & 0 & 17 \\
16 & Line Chart & 100.0 & 18 & 0 & 18 \\
17 & Parallel Coordinates & 100.0 & 18 & 0 & 18 \\
18 & Stacked Area & 94.4 & 17 & 1 & 18 \\
19 & Radar Chart & 100.0 & 13 & 0 & 13 \\
20 & Mosaic Chart & 100.0 & 13 & 0 & 13 \\
\midrule
\multicolumn{2}{l}{\textbf{Overall}} & \textbf{94.8} & \textbf{331} & \textbf{18} & \textbf{349} \\
\bottomrule
\end{tabular}
\end{table}

\subsection{Composite Visualizations}\label{sec: composite_evaluation}
Unlike basic charts, which are typically modeled as a single container with one visual mark type, composite visualizations are represented by multiple hierarchical containers with heterogeneous mark types. 
Different container hierarchies can produce visually similar compositions, making it difficult to match container structures for automatic evaluation.
Therefore, we invited 16 visualization practitioners (described in Sec.~\ref{sec: participant}) to manually evaluate the 20 composite visualizations. 
For each visualization, participants were shown the input image and the generated result and asked to answer three binary (\textit{Yes}/\textit{No}) questions:
By referring to the original visualization image,
does the generated visualization include all types of visual marks (Q1), reflect the correct composition pattern (Q2), and use the correct visual encodings (Q3)?
Q1 evaluates whether any visual mark types are missing, while Q2 assesses whether the generated visual components follow similar composition patterns to the input image, such as juxtaposition or superimposition. Q3 examines whether data are encoded using appropriate visual channels (e.g., position and color).
Since our DSL uses mocked data for visualization generation, participants were instructed to ignore differences in data distributions, such as clustering patterns in scatter plots, and focus solely on structural and encoding correctness.

\textbf{Results:}
Table~\ref{tab:evaluation_results_1} shows the evaluation results of composite visualization reproduction across 20 examples, indicating that \techName{} performs well in reproducing composite visualizations.
\begin{itemize}
    \item For \textit{\textbf{visual mark types}} (Q1), 15 cases successfully included all visual mark types. Three cases (\#05, \#06, \#08) received a small number of \textit{No} responses (fewer than four), which we attribute to the omission of auxiliary visual elements, such as dashed lines (\#06), point marks (\#05), and customized icons like arrows (\#08). 
    Two cases (\#14, \#20) received a large number of \textit{No} responses due to missing major visual marks, including area marks (\#14) and polylines (\#20). 
    Upon inspecting the corresponding DSL specifications, we found that containers for these marks were correctly generated, but certain attributes, such as control point layout for link marks, were incorrectly specified, preventing the marks from being rendered in the final visualization.
    \item For \textit{\textbf{composition}} (Q2), negative responses were mainly from data dependencies between containers (\#04), incorrect data structures (\#08), and missing spatial gaps between containers (\#16, \#20). In particular, our method currently generates mocked data independently for each container. As a result, in visualizations such as a line chart with dot marks (\#04), the line and dots may be generated from different data, causing the dots to deviate from the line. This issue can be addressed by assigning the same data to dependent containers using our interface.
    \item For \textit{\textbf{visual encoding}} (Q3), the primary issues arose from uniform color encodings caused by the use of template containers (\#07, \#17, \#18). 
    To enable data-driven generation of repetitive components, our DSL merges containers with identical specifications into a template container, and all instances of the template initially share the same mocked data. Thus, repeated components appear visually identical, particularly in color, but these color distributions become apparent after users upload their own data through the interface.
\end{itemize}


Overall, while reproduction of composite visualizations is not perfect, the observed issues can be effectively addressed through DSL editing and interactive refinement in \techName{}.

\begin{table}[htbp]
\centering
\caption{Evaluation Results of Composite Visualizations}
\label{tab:evaluation_results_1}
\footnotesize
\begin{tabular}{@{}clcccc@{}}
\toprule
\textbf{\#} & \textbf{Visualization} & \textbf{Mark\_Type} & \textbf{Composition} & \textbf{Encoding} \\
\midrule
01 & Faceted Bar Chart & 16/16 & 16/16 & 16/16 \\
02 & Faceted Scatter Plot & 16/16 & 15/16 & 16/16  \\
03 & Marginal Histograms & 16/16 & 16/16 & 16/16  \\
04 & Line with Highlights & 16/16 & 12/16 & 13/16  \\
05 & Box Plot & 14/16 & 15/16 & 10/16 \\
06 & OpinionSeer~\cite{wu2010opinionseer} & 12/16 & 13/16 & 13/16  \\
07 & iForest~\cite{zhao2018iforest} & 16/16 & 15/16 & 7/16  \\
08 & BitExtract~\cite{yue2018bitextract} & 13/16 & 14/16 & 14/16  \\
09 & DropoutSeer~\cite{chen2016dropoutseer} & 16/16 & 16/16 & 16/16  \\
10 & Clouddet~\cite{xu2019clouddet} & 16/16 & 16/16 & 16/16 \\
11 & Node Link Diagram & 16/16 & 16/16 & 16/16 \\
12 & Diverging Stacked Bar & 16/16 & 16/16 & 16/16  \\
13 & Line with Dots & 16/16 & 15/16 & 16/16 \\
14 & Line with Area & 4/16 & 8/16 & 8/16  \\
15 & Multiple Bar Charts & 16/16 & 16/16 & 16/16  \\
16 & Multiple Stacked Bar & 16/16 & 15/16 & 16/16  \\
17 & Multiple Area Charts & 16/16 & 16/16 & 8/16  \\
18 & LineUp~\cite{gratzl2013lineup} & 16/16 & 16/16 & 6/16  \\
19 & NameClarifier~\cite{shen2016nameclarifier} & 16/16 & 16/16 & 15/16 \\
20 & EnsembleLens~\cite{xu2018ensemblelens} & 7/16 & 13/16 & 13/16 \\ 
\midrule
\multicolumn{2}{l}{\textbf{Overall}} & \textbf{90.6\%} & \textbf{92.8\%} & \textbf{83.4\%} \\
\bottomrule
\end{tabular}
\end{table}

\section{User Interview}
This section describes the participants, interview procedures, and result analysis of the user interview.

\subsection{Participants and Apparatus} \label{sec: participant}
We recruited 16 visualization practitioners (U1–U16) from academia and industry, including researchers, designers, data analysts, and students. The participants (11 male, 5 female) were aged 20–34 and had varying levels of experience in visualization design and creation. Participants U1–U8 had more than three years of experience working with complex visualizations, such as designing composite visualizations in their daily work. In contrast, participants U9–U16 were considered novice practitioners, with between one and three years of experience and primarily creating basic charts, such as bar charts and scatter plots.
All interviews were held online via Zoom.
\techName{} was launched on the server, and participants can access it online via their laptops.
Each participant received compensation of \$15 per hour for their time.

\subsection{Procedures}
The interview consisted of three phases: tutorial, task, and interview. The tutorial phase introduced the background, DSL design, and \techName{}. One usage scenario was shown to illustrate how \techName{} supports reusing an image-based visualization. Participants then freely explored this example to familiarize themselves with \techName{}. The tutorial usually lasted 10 minutes.

In the task phase, participants are asked to examine 20 composite visualizations automatically generated by our pipeline (Sec.~\ref{sec: composite_evaluation}).
They then explored one of the visualizations in Sec.~\ref {sec: case}, which was not used in the tutorial.
Following a series of instructions, participants restructured the selected visualization, modified visual encodings, and updated the underlying data to explore the core functions of \techName{}. The instructions were intended for guidance rather than prescribing specific tasks, such as ``Please try to modify the composite structure.''
The task phase followed a think-aloud protocol, and participants were allowed to ask questions throughout the session.
The task phase continued until participants completed all instructions, typically within 40 minutes.

\begin{figure*}[!t]
    \centering
    \includegraphics[width=0.9\textwidth]{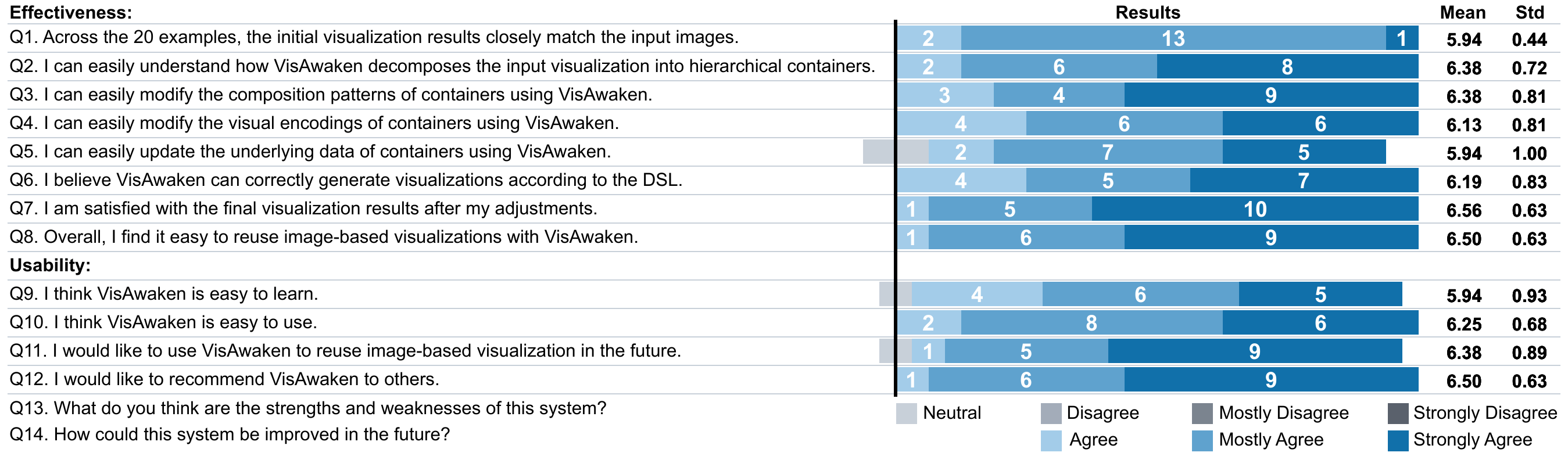}
    \caption{The user interview questionnaire results. Q1-Q12 are closed-ended questions rated on a 7-point Likert scale. Q13, Q14 are open-ended questions to collect participants’ feedback. The detailed scores of Q1-Q12 are shown in a stacked bar chart.
    }
    \label{fig: interview}
\end{figure*}

Finally, participants completed a post-study questionnaire (Fig.~\ref{fig: interview}) with 14 questions (Q1–Q14). Q1–Q12 were closed-ended questions rated on a 7-point Likert scale (strongly disagree to strongly agree)~\cite{joshi2015likert}, evaluating the effectiveness (Q1–Q8) and usability (Q9–Q12). Q13, Q14 were open-ended questions, gathering qualitative feedback on strengths, weaknesses, and potential improvements.
Overall, the entire interview lasted about one hour, and all the data collected from the participants were recorded with their permission.

\subsection{Results}
Fig.~\ref{fig: interview} shows the results of the questionnaire (Q1–Q12), including the score distribution, mean (M), and standard deviation (SD).
Overall, participants reported positive perceptions of \techName{} in reproducing, editing, and reusing image-based visualizations. A detailed analysis is presented below.

\textbf{Effectiveness:}
Participants reported that the automatically reproduced visualizations matched the input images in most cases (Q1, M=5.94). They also noted that only minimal DSL editing was required on mismatched attributes, suggesting that our pipeline can accurately reproduce image-based visualizations.
\techName{} also helped understand the process of decomposition (Q2, M=6.38).
Regarding editing and reuse, participants agreed that they could easily modify the DSL to change composite patterns (Q3, M=6.38), adjust visual encodings (Q4, M=6.13), and update underlying data (Q5, M=5.94). Two participants (U2, U11) reported neutral ratings for Q5, noting that \techName{} still requires manual data replacement for individual containers. They suggested more automated support for data processing and container-level data replacement.
Participants also expressed confidence in \techName{}’s ability to correctly generate visualizations according to the DSL specification (Q6, M=6.19). After completing their design adjustments, participants reported high satisfaction with the final visualization results (Q7, M=6.56). Finally, participants agreed that \techName{} made it easy to reuse image-based visualizations overall (Q8, M=6.50).

\textbf{Usability:}
Participants agreed that \techName{} was easy to learn (Q9, M=5.94) and easy to use (Q10, M=6.25), despite limited prior familiarity with the DSL paradigm. However, novice practitioners (U9–U16) reported slightly lower scores on Q9 and Q10 than expert practitioners (U1–U8), noting that the DSL was more challenging when attempting complex redesigns, such as creating new visual components not present in the original image.
Despite this learning curve, participants expressed intention to continue using \techName{} for reusing image-based visualizations (Q11, M=6.38) and a high willingness to recommend the system to others (Q12, M=6.50).

\textbf{Limitations and Suggestions:}
Participants phrased the ability to transform the static images into editable, data-driven designs and the template containers for simplifying design reuse, and the DSL-based interactions that provide fine-grained control of visual structure and encodings.
Regarding weaknesses, participants noted that while simple modifications could be easily achieved through DSL editing, the large combination space of specifications made it difficult to realize more complex redesign intentions. They also acknowledged that this complexity is necessary to support a wide range of visualizations.
For improvements, U11 suggested providing smoother onboarding support, while U2 and U9 proposed incorporating an LLM- or agent-based chatbot to capture user intent and automatically perform edits. 
U5 suggested enabling direct manipulation of visual elements instead of DSL editing, such as dragging bounding boxes to adjust coordinates. 
Additionally, three participants (U6, U7, U13) observed that the automatic generation was not always perfect and recommended a human–AI collaborative approach to DSL modification, for example, allowing users to brush a region and issue instructions to an LLM to refine the corresponding DSL.
Four participants (U1, U3, U7, U10) reported that data adaptation could be further improved. Although \techName{} provides a reference format, updating data still requires appropriate preprocessing, and containers need to be updated individually. In real-world applications, multiple visual components often share underlying data relationships or originate from the same data scope. Participants suggested supporting more effective data preparation across containers.
We appreciate the valuable feedback from participants and plan to incorporate data-dependency–aware data uploading, LLM-assisted modification, and direct visualization manipulation in future work.

\section{Discussion}
This section discusses the limitations of our approach and potential generalization directions for \techName{}.

\textbf{Limitations.}
\techName{} is not without limitations. 
Generating the DSL from visualization images using MLLMs remains a challenging task. Image resolution, visual clutter, and decorative elements (e.g., highlighting boxes) can negatively affect parsing accuracy. Visualizations with many components (e.g., repetitive glyphs) may also produce lengthy DSLs, increasing the model’s reasoning cost and occasionally leading to generation timeouts.
In addition, some visualizations rely on computational layouts and complex relations, often derived from well-processed underlying data, that are difficult to infer from images alone, such as treemaps, Sankey diagrams, force-directed layouts, and complex glyphs.
Data dependencies across containers are also not described in the DSL.
For example, in a line chart with overlaid point marks, \techName{} may treat the line and points as separate containers and mock their data independently, causing misalignment between the two. 

\textbf{Towards Flexible Visualization Reuse.}
Although image-based reproduction in \techName{} may be constrained by the capabilities of MLLMs, the DSL and container abstraction provide a promising foundation for more flexible visualization reuse. By representing designs as modular containers that independently render chart-type–agnostic components, designers can recombine and transform elements to create new composite visualizations~\cite{tsandilas2020structgraphics}.
Regarding basic composite patterns~\cite{javed2012exploring}, juxtaposition can be achieved by aligning containers along shared axes, overlays can be formed through overlapping coordinate regions, and nested or repetitive structures can be expressed through template containers that encapsulate multiple instances of a pattern. This supports the assembly of composite visualizations at the visual mark level, analogous to constructing dashboards from basic charts~\cite{chen2020composition}, but with finer-grained control over visual structure and composition.
Moreover, by abstracting away domain-specific data semantics, the container-based representation facilitates the comparison and migration of visualization design patterns across application domains. Collecting design images from diverse sources and extracting their container-based representations further enables the construction of reusable design corpora or galleries, supporting broader visualization reuse and redesign~\cite{chen2025visanatomysvgchartcorpus,bako2022understanding}.

\textbf{Towards AI for Composite Visualizations.}
Our DSL can express a wide range of composite visualizations and be directly converted into visualization images, enabling the systematic synthesis of complex visualization designs. Existing datasets for visualization image understanding primarily focus on basic chart types, such as bar and line charts, and largely overlook composite visualizations. By defining compositional rules over containers, layouts, and template patterns, diverse composite visualization DSLs can be automatically generated and rendered into images. This process supports the construction of large-scale synthetic datasets for complex visualization understanding and benchmarking, complementing existing chart-centric datasets for visualization reasoning and QA~\cite{wu2024chartinsights,masry2022chartqa,hu2019viznet}.
Furthermore, by mapping established composition strategies and design heuristics~\cite{javed2012exploring,munzner2014visualization} to DSL constraints, AI models may explore suitable composite visualizations based on data semantics and analytical goals. This direction extends prior work on visualization recommendation~\cite{hu2019vizml,li2021kg4vis}, dashboard generation~\cite{deng2022dashbot,lin2023dminer}, and design exploration~\cite{liu2025glyphweaver}.


\section{Conclusion and Future Work}
This work aims to enable reusable bitmap-image-based visualizations.
Specifically, we first introduce a generic DSL for decomposing and reproducing complex visualizations. Then, we propose \techName{}, a human-AI collaboration approach that enables users to reproduce, redesign, and reuse bitmap-image-based visualizations flexibly.
\techName{} employs an MLLM-based pipeline that automatically reproduces image-based visualizations based on the DSL and includes an interactive interface that supports result preview, DSL modification, and data adaptation.
In future work, we plan to improve the usability of \techName{} by incorporating an LLM-based chatbot and enabling AI-assisted data dispatching to reduce the learning curve and further ease redesign and reuse. 




\bibliographystyle{IEEEtran}
\bibliography{source/main}

@inproceedings{WangZWLW23,
  title        = {LLM4Vis: Explainable Visualization Recommendation using ChatGPT},
  author       = {Lei Wang and Songheng Zhang and Yun Wang and Ee-Peng Lim and Yong Wang},
  booktitle    = {Proceedings of the 2023 Conference on Empirical Methods in Natural Language Processing: Industry Track},
  editor       = {Mingxuan Wang and Imed Zitouni},
  pages        = {675--692},
  year         = {2023},
  doi          = {10.18653/v1/2023.emnlp-industry.64},
}

@book{munzner2014visualization,
  title={Visualization Analysis and Design},
  author={Munzner, Tamara},
  year={2014},
  publisher={CRC Press}
}

@book{fowler2010domain,
  title={Domain-specific languages},
  author={Fowler, Martin},
  year={2010},
  publisher={Pearson Education}
}

@article{stolte2002polaris,
	title        = {Polaris: A System for Query, Analysis, and Visualization of Multidimensional Relational Databases},
	author       = {Stolte, Chris and Tang, Diane and Hanrahan, Pat},
	year         = 2002,
	journal      = {IEEE Transactions on Visualization and Computer Graphics},
	volume       = 8,
	number       = 1,
	pages        = {52--65}
}

@inproceedings{liu2025simvecvis,
  author={Liu, Can and Da, Chunlin and Long, Xiaoxiao and Yang, Yuxiao and Zhang, Yu and Wang, Yong},
  booktitle={Proceedings of the 2025 IEEE VIS Conference}, 
  title={SimVecVis: A Dataset for Enhancing MLLMs in Visualization Understanding}, 
  year={2025},
  volume={},
  number={},
  pages={26-30},
  doi={10.1109/VIS60296.2025.00010}}

@inproceedings{hoffswell2018setcola,
  title={Setcola: High-level constraints for graph layout},
  author={Hoffswell, Jane and Borning, Alan and Heer, Jeffrey},
  booktitle={Computer Graphics Forum},
  volume={37},
  number={3},
  pages={537--548},
  year={2018},
  organization={Wiley Online Library}
}

@inproceedings{liu2021atlas,
  title={Atlas: Grammar-based procedural generation of data visualizations},
  author={Liu, Zhicheng and Chen, Chen and Morales, Francisco and Zhao, Yishan},
  booktitle={Proceedings of IEEE Visualization Conference (VIS)},
  pages={171--175},
  year={2021},
  organization={IEEE}
}

@article{park2017atom,
  title={Atom: A grammar for unit visualizations},
  author={Park, Deokgun and Drucker, Steven M and Fernandez, Roland and Elmqvist, Niklas},
  journal={IEEE Transactions on Visualization and Computer Graphics},
  volume={24},
  number={12},
  pages={3032--3043},
  year={2017},
  publisher={IEEE}
}

@ARTICLE{Bako2023understanding,
  author={Bako, Hannah K. and Liu, Xinyi and Battle, Leilani and Liu, Zhicheng},
  journal={IEEE Transactions on Visualization and Computer Graphics}, 
  title={Understanding how Designers Find and Use Data Visualization Examples}, 
  year={2023},
  volume={29},
  number={1},
  pages={1048-1058}
}

@inproceedings{DeconstructingD3,
author = {Harper, Jonathan and Agrawala, Maneesh},
title = {Deconstructing and restyling D3 visualizations},
year = {2014},
isbn = {9781450330695},
booktitle = {Proceedings of the 27th Annual ACM Symposium on User Interface Software and Technology},
pages = {253–262},
numpages = {10},
location = {Honolulu, Hawaii, USA},
}

@ARTICLE{deng2023revisit,
  author={Deng, Dazhen and Cui, Weiwei and Meng, Xiyu and Xu, Mengye and Liao, Yu and Zhang, Haidong and Wu, Yingcai},
  journal={IEEE Transactions on Visualization and Computer Graphics}, 
  title={Revisiting the Design Patterns of Composite Visualizations}, 
  year={2023},
  volume={29},
  number={12},
  pages={5406-5421},
  doi={10.1109/TVCG.2022.3213565}}

@article{xie2025datawink,
  title={DataWink: Reusing and Adapting SVG-based Visualization Examples with Large Multimodal Models},
  author={Xie, Liwenhan and Lin, Yanna and Liu, Can and Qu, Huamin and Shu, Xinhuan},
  journal={arXiv preprint arXiv:2507.17734},
  year={2025}
}

@inproceedings{li2022structure,
  title={Structure-aware visualization retrieval},
  author={Li, Haotian and Wang, Yong and Wu, Aoyu and Wei, Huan and Qu, Huamin},
  booktitle={Proceedings of the 2022 CHI Conference on Human Factors in Computing Systems},
  pages={1--14},
  year={2022}
}

@article{chen2023mystique,
  title={Mystique: Deconstructing svg charts for layout reuse},
  author={Chen, Chen and Lee, Bongshin and Wang, Yunhai and Chang, Yunjeong and Liu, Zhicheng},
  journal={IEEE Transactions on Visualization and Computer Graphics},
  volume={30},
  number={1},
  pages={447--457},
  year={2023},
  publisher={IEEE}
}

@article{harper2017converting,
  title={Converting basic D3 charts into reusable style templates},
  author={Harper, Jonathan and Agrawala, Maneesh},
  journal={IEEE Transactions on Visualization and Computer Graphics},
  volume={24},
  number={3},
  pages={1274--1286},
  year={2017},
  publisher={IEEE}
}

@article{satyanarayan2016vega,
  title={Vega-lite: A grammar of interactive graphics},
  author={Satyanarayan, Arvind and Moritz, Dominik and Wongsuphasawat, Kanit and Heer, Jeffrey},
  journal={IEEE Transactions on Visualization and Computer Graphics},
  volume={23},
  number={1},
  pages={341--350},
  year={2016},
  publisher={IEEE}
}

@inproceedings{poco2017reverse,
  title={Reverse-engineering visualizations: Recovering visual encodings from chart images},
  author={Poco, Jorge and Heer, Jeffrey},
  booktitle={Computer Graphics Forum},
  volume={36},
  number={3},
  pages={353--363},
  year={2017},
  organization={Wiley Online Library}
}

@incollection{wilkinson2011grammar,
  title={The grammar of graphics},
  author={Wilkinson, Leland},
  booktitle={Handbook of computational statistics: Concepts and methods},
  pages={375--414},
  year={2011},
  publisher={Springer}
}

@article{ying2024reviving,
  author={Ying, Lu and Wang, Yun and Li, Haotian and Dou, Shuguang and Zhang, Haidong and Jiang, Xinyang and Qu, Huamin and Wu, Yingcai},
  journal={IEEE Transactions on Visualization and Computer Graphics}, 
  title={Reviving Static Charts Into Live Charts}, 
  year={2025},
  volume={31},
  number={8},
  pages={4314-4328},
  doi={10.1109/TVCG.2024.3397004}}

@inproceedings{savva2011revision,
  title={Revision: Automated classification, analysis and redesign of chart images},
  author={Savva, Manolis and Kong, Nicholas and Chhajta, Arti and Fei-Fei, Li and Agrawala, Maneesh and Heer, Jeffrey},
  booktitle={Proceedings of the 24th annual ACM Symposium on User Interface Software and Technology},
  pages={393--402},
  year={2011}
}

@inproceedings{jung2017chartsense,
  title={Chartsense: Interactive data extraction from chart images},
  author={Jung, Daekyoung and Kim, Wonjae and Song, Hyunjoo and Hwang, Jeong-in and Lee, Bongshin and Kim, Bohyoung and Seo, Jinwook},
  booktitle={Proceedings of the 2017 CHI Conference on Human Factors in Computing Systems},
  pages={6706--6717},
  year={2017}
}

@article{jo2018declarative,
  title={A declarative rendering model for multiclass density maps},
  author={Jo, Jaemin and Vernier, Fr{\'e}d{\'e}ric and Dragicevic, Pierre and Fekete, Jean-Daniel},
  journal={IEEE Transactions on Visualization and Computer Graphics},
  volume={25},
  number={1},
  pages={470--480},
  year={2018},
  publisher={IEEE}
}

@ARTICLE{choi2014vivaldi,
  author={Choi, Hyungsuk and Choi, Woohyuk and Quan, Tran Minh and Hildebrand, David G. C. and Pfister, Hanspeter and Jeong, Won-Ki},
  journal={IEEE Transactions on Visualization and Computer Graphics}, 
  title={Vivaldi: A Domain-Specific Language for Volume Processing and Visualization on Distributed Heterogeneous Systems}, 
  year={2014},
  volume={20},
  number={12},
  pages={2407-2416},
  doi={10.1109/TVCG.2014.2346322}}

@ARTICLE{li2023contour,
  author={Li, Sihang and Yu, Jiacheng and Li, Mingxuan and Liu, Le and Zhang, Xiaolong Luke and Yuan, Xiaoru},
  journal={IEEE Transactions on Visualization and Computer Graphics}, 
  title={A Framework for Multiclass Contour Visualization}, 
  year={2023},
  volume={29},
  number={1},
  pages={353-362},
  doi={10.1109/TVCG.2022.3209482}}

@article{liu2024manipulable,
  author={Liu, Zhicheng and Chen, Chen and Hooker, John},
  journal={IEEE Transactions on Visualization and Computer Graphics}, 
  title={Manipulable Semantic Components: A Computational Representation of Data Visualization Scenes}, 
  year={2025},
  volume={31},
  number={1},
  pages={732-742},
  doi={10.1109/TVCG.2024.3456296}}

@ARTICLE{9585700,
  author={Cui, Weiwei and Wang, Jinpeng and Huang, He and Wang, Yun and Lin, Chin-Yew and Zhang, Haidong and Zhang, Dongmei},
  journal={IEEE Transactions on Visualization and Computer Graphics}, 
  title={A Mixed-Initiative Approach to Reusing Infographic Charts}, 
  year={2022},
  volume={28},
  number={1},
  pages={173-183},
  doi={10.1109/TVCG.2021.3114856}}

@inproceedings{pollock2024bluefish,
  title={Bluefish: Composing diagrams with declarative relations},
  author={Pollock, Josh and Mei, Catherine and Huang, Grace and Evans, Elliot and Jackson, Daniel and Satyanarayan, Arvind},
  booktitle={Proceedings of the 37th Annual ACM Symposium on User Interface Software and Technology},
  pages={1--21},
  year={2024}
}

@article{tao2020kyrixs,
 author={Tao, Wenbo and Hou, Xinli and Sah, Adam and Battle, Leilani and Chang, Remco and Stonebraker, Michael},
  journal={IEEE Transactions on Visualization and Computer Graphics}, 
  title={Kyrix-S: Authoring Scalable Scatterplot Visualizations of Big Data}, 
  year={2021},
  volume={27},
  number={2},
  pages={401-411},
  doi={10.1109/TVCG.2020.3030372}}

@article{hong2025llms,
  author={Hong, Jiayi and Seto, Christian and Fan, Arlen and Maciejewski, Ross},
  journal={IEEE Transactions on Visualization and Computer Graphics}, 
  title={Do LLMs Have Visualization Literacy? An Evaluation on Modified Visualizations to Test Generalization in Data Interpretation}, 
  year={2025},
  volume={31},
  number={10},
  pages={7004-7018},
  doi={10.1109/TVCG.2025.3536358}}

@article{das2025charts,
  title={Charts-of-Thought: Enhancing LLM Visualization Literacy Through Structured Data Extraction},
  author={Das, Amit Kumar and Tarun, Mohammad and Mueller, Klaus},
  journal={arXiv preprint arXiv:2508.04842},
  year={2025}
}

@article{li2024visualization,
  title={Visualization generation with large language models: An evaluation},
  author={Wang, Xinyu and Liang, Chenwei and Zheng, Shunyuan and Liang, Jinyuan and Li, Guozheng and Zhang, Yu and Liu, Chi Harold},
  journal={arXiv preprint arXiv:2401.11255},
  year={2024}
}

@article{chen2024viseval,
  author={Chen, Nan and Zhang, Yuge and Xu, Jiahang and Ren, Kan and Yang, Yuqing},
  journal={IEEE Transactions on Visualization and Computer Graphics}, 
  title={VisEval: A Benchmark for Data Visualization in the Era of Large Language Models}, 
  year={2025},
  volume={31},
  number={1},
  pages={1301-1311},
  doi={10.1109/TVCG.2024.3456320}}

@article{xie2025datasway,
       title={DataSway: Vivifying Metaphoric Visualization with Animation Clip Generation and Coordination},
  author={Xie, Liwenhan and Zhou, Jiayi and Rao, Anyi and Qu, Huamin and Shu, Xinhuan},
  journal={arXiv preprint arXiv:2507.22051},
  year={2025}
}

@ARTICLE{deng2023composite,
  author={Deng, Dazhen and Cui, Weiwei and Meng, Xiyu and Xu, Mengye and Liao, Yu and Zhang, Haidong and Wu, Yingcai},
  journal={IEEE Transactions on Visualization and Computer Graphics}, 
  title={Revisiting the Design Patterns of Composite Visualizations}, 
  year={2023},
  volume={29},
  number={12},
  pages={5406-5421},
  doi={10.1109/TVCG.2022.3213565}}

@inproceedings{li2020gotree,
author = {Li, Guozheng and Tian, Min and Xu, Qinmei and McGuffin, Michael J. and Yuan, Xiaoru},
title = {GoTree: A Grammar of Tree Visualizations},
year = {2020},
isbn = {9781450367080},
publisher = {ACM},
booktitle = {Proceedings of the 2020 CHI Conference on Human Factors in Computing Systems},
pages = {1–13},
numpages = {13},
location = {Honolulu, HI, USA},
}

@article{chen2025visanatomysvgchartcorpus,
  author={Chen, Chen and Bako, Hannah K. and Yu, Peihong and Hooker, John and Joyal, Jeffrey and Wang, Simon C. and Kim, Samuel and Wu, Jessica and Ding, Aoxue and Sandeep, Lara and Chen, Alex and Sinha, Chayanika and Liu, Zhicheng},
  journal={IEEE Transactions on Visualization and Computer Graphics}, 
  title={VISANATOMY: an SVG Chart Corpus with Fine-Grained Semantic Labels}, 
  year={2025},
  volume={},
  number={},
  pages={1-11},
  doi={10.1109/TVCG.2025.3634263}}

@inproceedings{warner2023interactive,
  title={Interactive flexible style transfer for vector graphics},
  author={Warner, Jeremy and Kim, Kyu Won and Hartmann, Bjoern},
  booktitle={Proceedings of the 36th Annual ACM Symposium on User Interface Software and Technology},
  pages={1--14},
  year={2023}
}

@inproceedings{kumar2011bricolage,
  title={Bricolage: example-based retargeting for web design},
  author={Kumar, Ranjitha and Talton, Jerry O and Ahmad, Salman and Klemmer, Scott R},
  booktitle={Proceedings of the SIGCHI Conference on Human Factors in Computing Systems},
  pages={2197--2206},
  year={2011}
}

@inproceedings{baigelenov2025visualization,
  title={How Visualization Designers Perceive and Use Inspiration},
  author={Baigelenov, Ali and Shukla, Prakash and Parsons, Paul},
  booktitle={Proceedings of the 2025 CHI Conference on Human Factors in Computing Systems},
  pages={1--13},
  year={2025}
}

@article{li2025metal,
  title={Metal: A multi-agent framework for chart generation with test-time scaling},
  author={Li, Bingxuan and Wang, Yiwei and Gu, Jiuxiang and Chang, Kai-Wei and Peng, Nanyun},
  journal={arXiv preprint arXiv:2502.17651},
  year={2025}
}

@article{zhu2019towards,
  title={Towards automated infographic design: Deep learning-based auto-extraction of extensible timeline},
  author={Zhu-Tian, Chen and Wang, Yun and Wang, Qianwen and Wang, Yong and Qu, Huamin},
  journal={IEEE Transactions on Visualization and Computer Graphics},
  volume={26},
  number={1},
  pages={917--926},
  year={2019},
  publisher={IEEE}
}

@article{yang2024chartmimic,
  title={Chartmimic: Evaluating lmm's cross-modal reasoning capability via chart-to-code generation},
  author={Yang, Cheng and Shi, Chufan and Liu, Yaxin and Shui, Bo and Wang, Junjie and Jing, Mohan and Xu, Linran and Zhu, Xinyu and Li, Siheng and Zhang, Yuxiang and others},
  journal={arXiv preprint arXiv:2406.09961},
  year={2024}
}

@article{joshi2015likert,
  title={Likert scale: Explored and explained},
  author={Joshi, Ankur and Kale, Saket and Chandel, Satish and Pal, D Kumar},
  journal={British journal of applied science \& technology},
  volume={7},
  number={4},
  pages={396},
  year={2015},
  publisher={Sciencedomain International}
}

@article{wu2010opinionseer,
  title={OpinionSeer: interactive visualization of hotel customer feedback},
  author={Wu, Yingcai and Wei, Furu and Liu, Shixia and Au, Norman and Cui, Weiwei and Zhou, Hong and Qu, Huamin},
  journal={IEEE Transactions on Visualization and Computer Graphics},
  volume={16},
  number={6},
  pages={1109--1118},
  year={2010},
  publisher={IEEE}
}

@article{zhao2018iforest,
  title={iforest: Interpreting random forests via visual analytics},
  author={Zhao, Xun and Wu, Yanhong and Lee, Dik Lun and Cui, Weiwei},
  journal={IEEE Transactions on Visualization and Computer Graphics},
  volume={25},
  number={1},
  pages={407--416},
  year={2018},
  publisher={IEEE}
}

@article{bostock2011d3,
  title={D$^3$ data-driven documents},
  author={Bostock, Michael and Ogievetsky, Vadim and Heer, Jeffrey},
  journal={IEEE Transactions on Visualization and Computer Graphics},
  volume={17},
  number={12},
  pages={2301--2309},
  year={2011},
  publisher={IEEE}
}

@inproceedings{javed2012exploring,
  title={Exploring the design space of composite visualization},
  author={Javed, Waqas and Elmqvist, Niklas},
  booktitle={Proceedings of 2012 IEEE Pacific Visualization Symposium},
  pages={1--8},
  year={2012},
  organization={IEEE}
}

@article{tsandilas2020structgraphics,
  title={StructGraphics: Flexible visualization design through data-agnostic and reusable graphical structures},
  author={Tsandilas, Theophanis},
  journal={IEEE Transactions on Visualization and Computer Graphics},
  volume={27},
  number={2},
  pages={315--325},
  year={2020},
  publisher={IEEE}
}

@article{chen2020composition,
  title={Composition and configuration patterns in multiple-view visualizations},
  author={Chen, Xi and Zeng, Wei and Lin, Yanna and Ai-Maneea, Hayder Mahdi and Roberts, Jonathan and Chang, Remco},
  journal={IEEE Transactions on Visualization and Computer Graphics},
  volume={27},
  number={2},
  pages={1514--1524},
  year={2020},
  publisher={IEEE}
}

@article{bako2022understanding,
  title={Understanding how designers find and use data visualization examples},
  author={Bako, Hannah K and Liu, Xinyi and Battle, Leilani and Liu, Zhicheng},
  journal={IEEE Transactions on Visualization and Computer Graphics},
  volume={29},
  number={1},
  pages={1048--1058},
  year={2022},
  publisher={IEEE}
}

@inproceedings{masry2022chartqa,
  title={Chartqa: A benchmark for question answering about charts with visual and logical reasoning},
  author={Masry, Ahmed and Do, Xuan Long and Tan, Jia Qing and Joty, Shafiq and Hoque, Enamul},
  booktitle={Findings of the Association for Computational Linguistics: ACL 2022},
  pages={2263--2279},
  year={2022}
}

@article{wu2024chartinsights,
  title={Chartinsights: Evaluating multimodal large language models for low-level chart question answering},
  author={Wu, Yifan and Yan, Lutao and Shen, Leixian and Wang, Yunhai and Tang, Nan and Luo, Yuyu},
  journal={arXiv preprint arXiv:2405.07001},
  year={2024}
}

@article{deng2022dashbot,
  title={DashBot: Insight-driven dashboard generation based on deep reinforcement learning},
  author={Deng, Dazhen and Wu, Aoyu and Qu, Huamin and Wu, Yingcai},
  journal={IEEE Transactions on Visualization and Computer Graphics},
  volume={29},
  number={1},
  pages={690--700},
  year={2022},
  publisher={IEEE}
}

@article{li2021kg4vis,
  title={KG4Vis: A knowledge graph-based approach for visualization recommendation},
  author={Li, Haotian and Wang, Yong and Zhang, Songheng and Song, Yangqiu and Qu, Huamin},
  journal={IEEE Transactions on Visualization and Computer Graphics},
  volume={28},
  number={1},
  pages={195--205},
  year={2021},
  publisher={IEEE}
}

@inproceedings{hu2019vizml,
  title={Vizml: A machine learning approach to visualization recommendation},
  author={Hu, Kevin and Bakker, Michiel A and Li, Stephen and Kraska, Tim and Hidalgo, C{\'e}sar},
  booktitle={Proceedings of the 2019 CHI Conference on Human Factors in Computing Systems},
  pages={1--12},
  year={2019}
}

@inproceedings{hu2019viznet,
  title={VizNet: Towards a large-scale visualization learning and benchmarking repository},
  author={Hu, Kevin and Gaikwad, Snehalkumar'Neil'S and Hulsebos, Madelon and Bakker, Michiel A and Zgraggen, Emanuel and Hidalgo, C{\'e}sar and Kraska, Tim and Li, Guoliang and Satyanarayan, Arvind and Demiralp, {\c{C}}a{\u{g}}atay},
  booktitle={Proceedings of the 2019 CHI Conference on Human Factors in Computing Systems},
  pages={1--12},
  year={2019}
}

@article{liu2025glyphweaver,
  title={GlyphWeaver: Unlocking Glyph Design Creativity with Uniform Glyph DSL and AI},
  author={Liu, Can and Chen, Shiwei and Jiang, Zhibang and Wang, Yong},
  journal={arXiv preprint arXiv:2509.08444},
  year={2025}
}

@article{lin2023dminer,
  title={DMiner: Dashboard design mining and recommendation},
  author={Lin, Yanna and Li, Haotian and Wu, Aoyu and Wang, Yong and Qu, Huamin},
  journal={IEEE Transactions on Visualization and Computer Graphics},
  volume={30},
  number={7},
  pages={4108--4121},
  year={2023},
  publisher={IEEE}
}

@article{yue2018bitextract,
  title={Bitextract: Interactive visualization for extracting bitcoin exchange intelligence},
  author={Yue, Xuanwu and Shu, Xinhuan and Zhu, Xinyu and Du, Xinnan and Yu, Zheqing and Papadopoulos, Dimitrios and Liu, Siyuan},
  journal={IEEE Transactions on Visualization and Computer Graphics},
  volume={25},
  number={1},
  pages={162--171},
  year={2018},
  publisher={IEEE}
}

@inproceedings{chen2016dropoutseer,
  title={DropoutSeer: Visualizing learning patterns in Massive Open Online Courses for dropout reasoning and prediction},
  author={Chen, Yuanzhe and Chen, Qing and Zhao, Mingqian and Boyer, Sebastien and Veeramachaneni, Kalyan and Qu, Huamin},
  booktitle={Proceedings of 2016 IEEE Conference on Visual Analytics Science and Technology},
  pages={111--120},
  year={2016},
  organization={IEEE}
}

@article{xu2019clouddet,
  title={Clouddet: Interactive visual analysis of anomalous performances in cloud computing systems},
  author={Xu, Ke and Wang, Yun and Yang, Leni and Wang, Yifang and Qiao, Bo and Qin, Si and Xu, Yong and Zhang, Haidong and Qu, Huamin},
  journal={IEEE Transactions on Visualization and Computer Graphics},
  volume={26},
  number={1},
  pages={1107--1117},
  year={2019},
  publisher={IEEE}
}

@article{gratzl2013lineup,
  title={Lineup: Visual analysis of multi-attribute rankings},
  author={Gratzl, Samuel and Lex, Alexander and Gehlenborg, Nils and Pfister, Hanspeter and Streit, Marc},
  journal={IEEE Transactions on Visualization and Computer Graphics},
  volume={19},
  number={12},
  pages={2277--2286},
  year={2013},
  publisher={IEEE}
}

@article{shen2016nameclarifier,
  title={Nameclarifier: A visual analytics system for author name disambiguation},
  author={Shen, Qiaomu and Wu, Tongshuang and Yang, Haiyan and Wu, Yanhong and Qu, Huamin and Cui, Weiwei},
  journal={IEEE Transactions on Visualization and Computer Graphics},
  volume={23},
  number={1},
  pages={141--150},
  year={2016},
  publisher={IEEE}
}

@article{xu2018ensemblelens,
  title={Ensemblelens: Ensemble-based visual exploration of anomaly detection algorithms with multidimensional data},
  author={Xu, Ke and Xia, Meng and Mu, Xing and Wang, Yun and Cao, Nan},
  journal={IEEE Transactions on Visualization and Computer Graphics},
  volume={25},
  number={1},
  pages={109--119},
  year={2018},
  publisher={IEEE}
}

@article{song2022vividgraph,
  title={VividGraph: Learning to extract and redesign network graphs from visualization images},
  author={Song, Sicheng and Li, Chenhui and Sun, Yujing and Wang, Changbo},
  journal={IEEE Transactions on Visualization and Computer Graphics},
  volume={29},
  number={7},
  pages={3169--3181},
  year={2022},
  publisher={IEEE}
}

@article{song2024gvvst,
  author={Song, Sicheng and Zhang, Yipeng and Lin, Yanna and Qu, Huamin and Wang, Changbo and Li, Chenhui},
  journal={IEEE Transactions on Visualization and Computer Graphics}, 
  title={GVVST: Image-Driven Style Extraction From Graph Visualizations for Visual Style Transfer}, 
  year={2025},
  volume={31},
  number={9},
  pages={5975-5989},
  doi={10.1109/TVCG.2024.3485701}}

@inproceedings{ying2024vaid,
  title={VAID: Indexing view designs in visual analytics system},
  author={Ying, Lu and Wu, Aoyu and Li, Haotian and Deng, Zikun and Lan, Ji and Wu, Jiang and Wang, Yong and Qu, Huamin and Deng, Dazhen and Wu, Yingcai},
  booktitle={Proceedings of the 2024 CHI Conference on Human Factors in Computing Systems},
  pages={1--15},
  year={2024}
}

@inproceedings{borgo2013glyph,
  author    = {Borgo, Rita and Kehrer, Johannes and Chung, David H. S. and Maguire, Eamonn and Laramee, Robert S. and Hauser, Helwig and Ward, Matthew and Chen, Min},
  title     = {Glyph-based Visualization: Foundations, Design Guidelines, Techniques and Applications},
  booktitle = {Proccedings of Eurographics (State of the Art Reports)},
  year      = {2013},
  pages     = {39--63}
}

@book{tufte1990envisioning,
  author    = {Tufte, Edward R. and Goeler, Nora Hillman and Benson, Richard},
  title     = {Envisioning Information},
  year      = {1990},
  publisher = {Graphics Press},
  address   = {Cheshire, CT},
  volume    = {126}
}


 
\vspace{-33pt}
\begin{IEEEbiography}[{\includegraphics[width=1in,height=1.25in,clip,keepaspectratio]{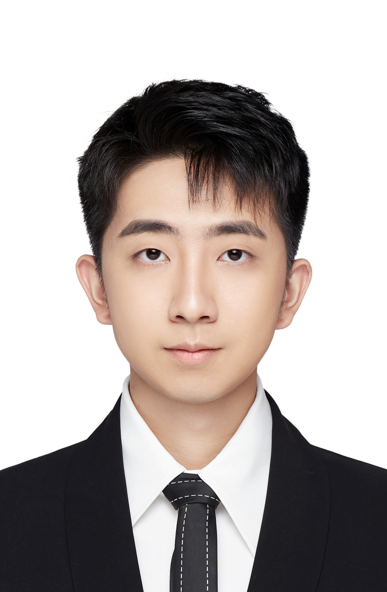}}]{Xiaolin Wen}
is currently a Ph.D student in the College of Computing and Data Science, Nanyang Technological University (NTU). 
His research interests mainly focus on visualization for FinTech and LLM-assisted design study.
He received his master's degree in Computer Science and Technology from Sichuan University in 2023 and his dual bachelor's degree in Computer Science and Financial Engineering from Sichuan University in 2016.
For more information, kindly visit \url{https://wenxiaolin.com/}.
\end{IEEEbiography}

\vspace{-33pt}

\begin{IEEEbiography}[{\includegraphics[width=1in,height=1.2in,clip,keepaspectratio]{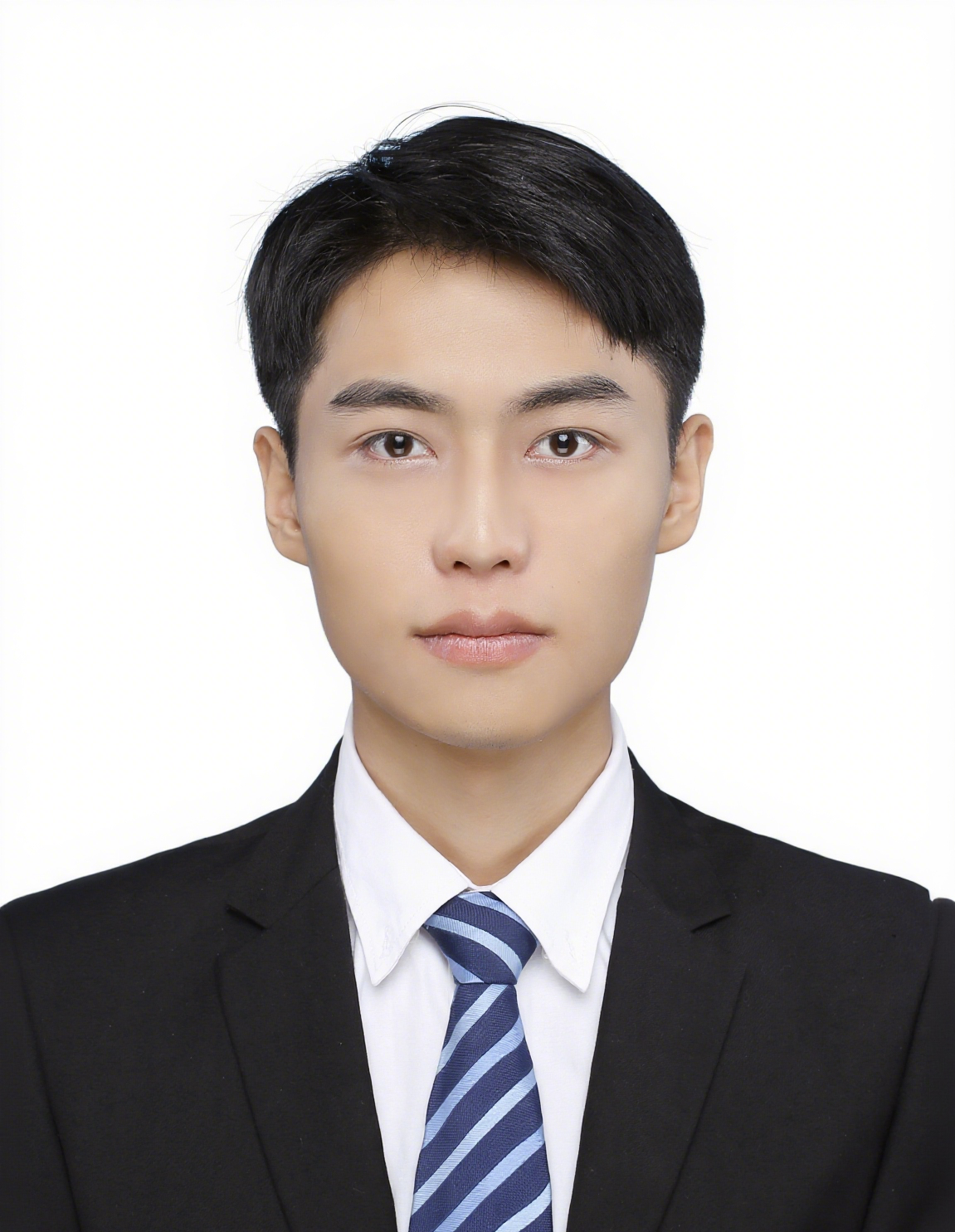}
}]{Changlin Li} is currently a front-end development engineer at a Chinese Internet technology company. His research interests mainly focus on visualization. He received his master's degree in Computer Science and Technology from Sichuan University in 2023.
\end{IEEEbiography}

\vspace{-33pt}

\begin{IEEEbiography}[{\includegraphics[width=1in,height=1.25in,clip,keepaspectratio]{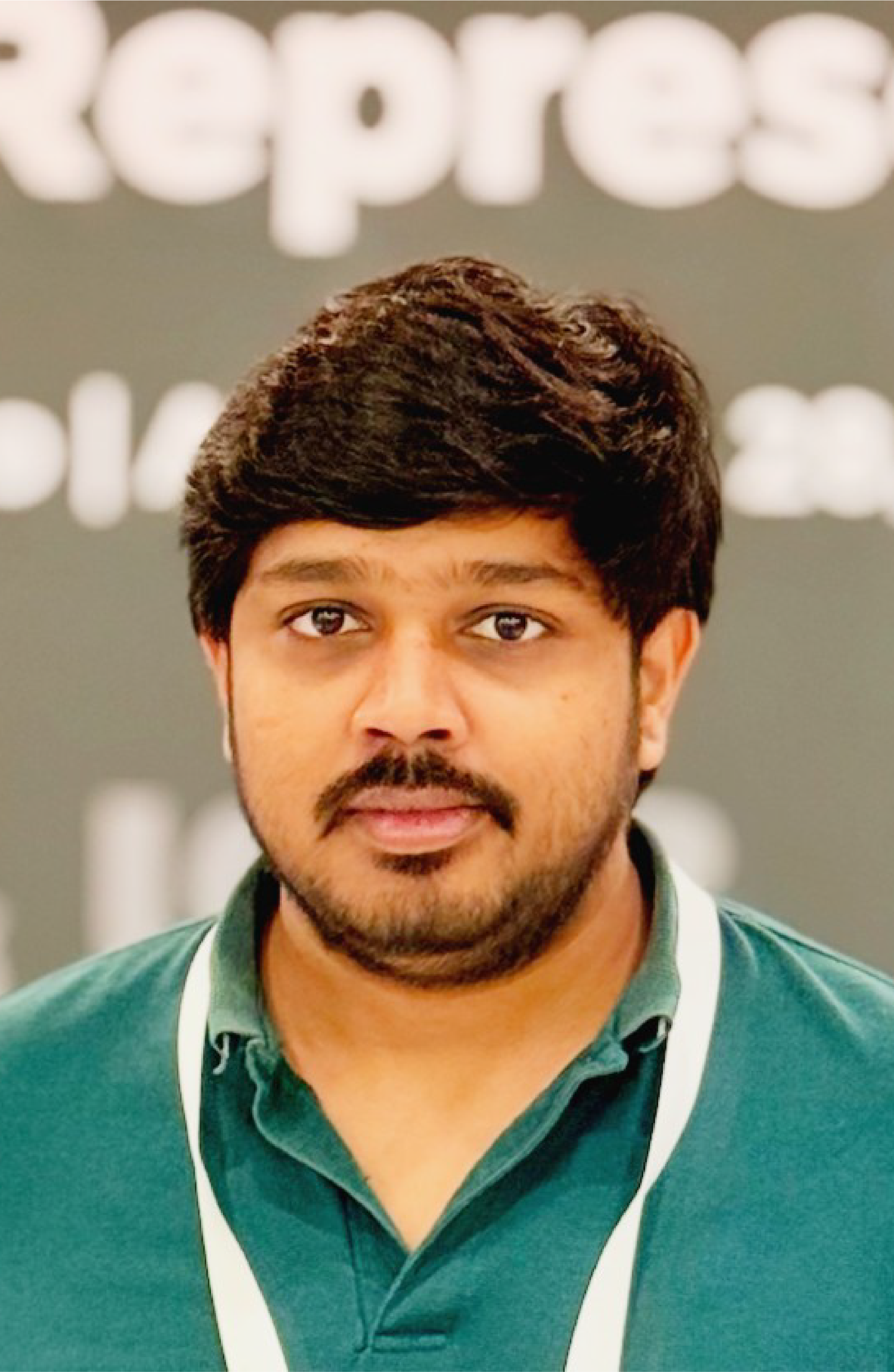}}]{Manusha Karunathilaka} is currently a Ph.D. candidate in School of Computing and Information Systems at Singapore Management University (SMU). His research interests include data visualization, storytelling, and human–computer interaction, with a particular focus on automated visual story generation. He received his bachelor’s degree in Computer Science and Engineering from the University of Moratuwa, Sri Lanka. For more details, kindly visit \url{https://manusha-karunathilaka.com}.
\end{IEEEbiography}

\vspace{-33pt}

\begin{IEEEbiography}[{\includegraphics[width=1in,height=1.25in,clip,keepaspectratio]{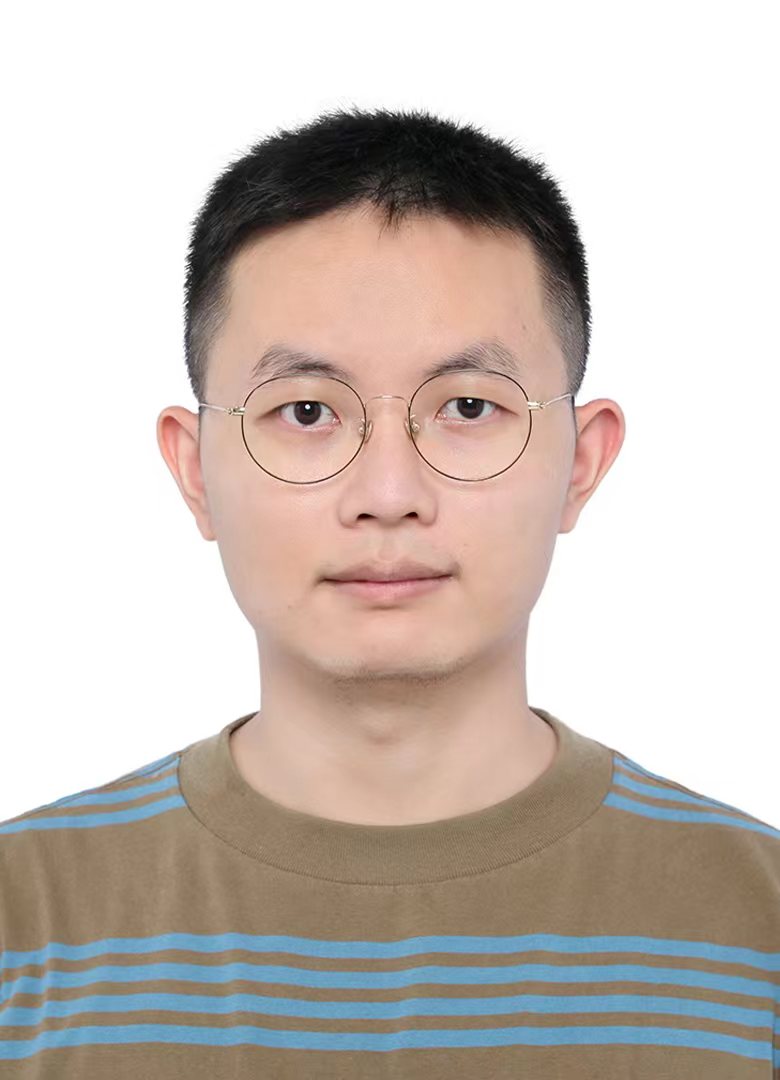}}]{Can Liu} is a research fellow in College of Computing and Data Science, Nanyang Technological University, Singapore.
He received his Ph.D. in Intelligent Science (2023), B.Sc. in Computer Science (2018), and B.Ec in Economics (2018) degrees from Peking University.
His research interests lie in the field of deep learning-driven visualization, especially intelligent interaction for visualization. For more details, please visit \url{https://liucan.me}.
\end{IEEEbiography}

\vspace{-33pt}

\begin{IEEEbiography}[{\includegraphics[width=1in,height=1.25in,clip,keepaspectratio]{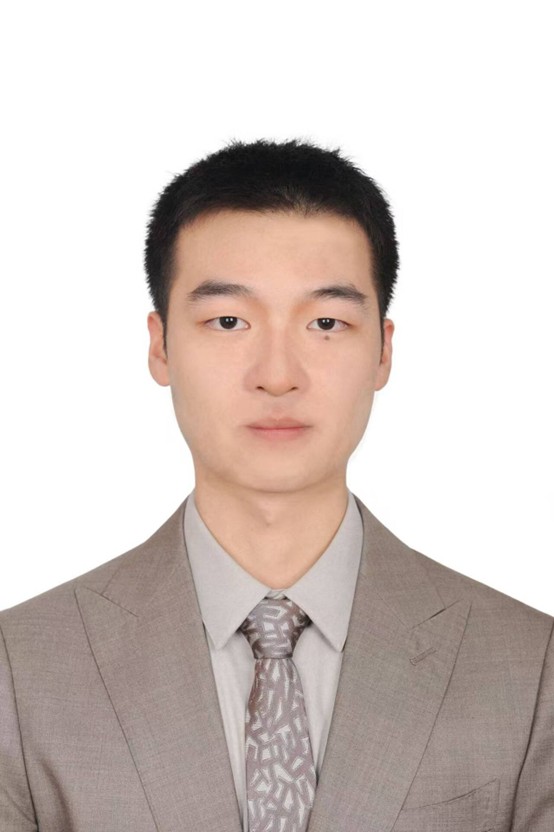}}]{Fangzhuo Jin}
is currently pursuing the B.E. degree in Software Engineering at Huazhong University of Science and Technology, China. His research interests include data visualization and human-AI collaboration.
\end{IEEEbiography}

\vspace{-3pt}

\begin{IEEEbiography}[{\includegraphics[width=1in,height=1.25in,clip,keepaspectratio]{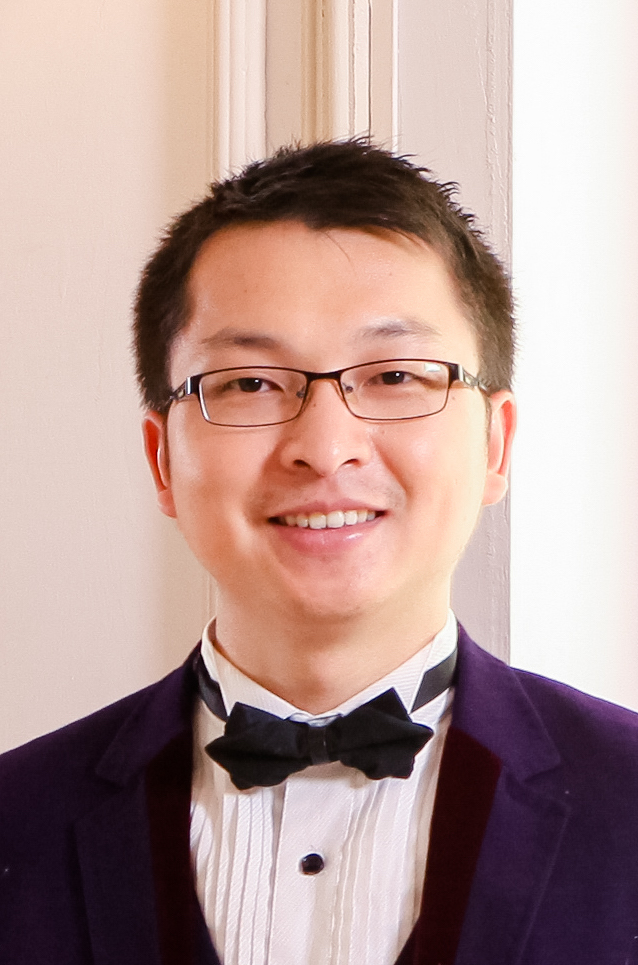}}]{Yong Wang} is currently an assistant professor in the College of Computing and Data Science, Nanyang Technological University. His research interests include data visualization, HCI, and human-AI collaboration, with an emphasis on their application to FinTech, quantum computing, and online learning. He obtained his Ph.D. in Computer Science from Hong Kong University of Science and Technology. He received his B.E. and M.E. from Harbin Institute of Technology and Huazhong University of Science and Technology, respectively. For more details, please refer to \url{http://yong-wang.org}.
\end{IEEEbiography}


\vfill

\clearpage
\onecolumn
\appendix
\section*{A. Composite Visualizations used for the Quantitative Evaluation}
\begin{figure}[H]
    \centering
    \includegraphics[width=0.96\textwidth]{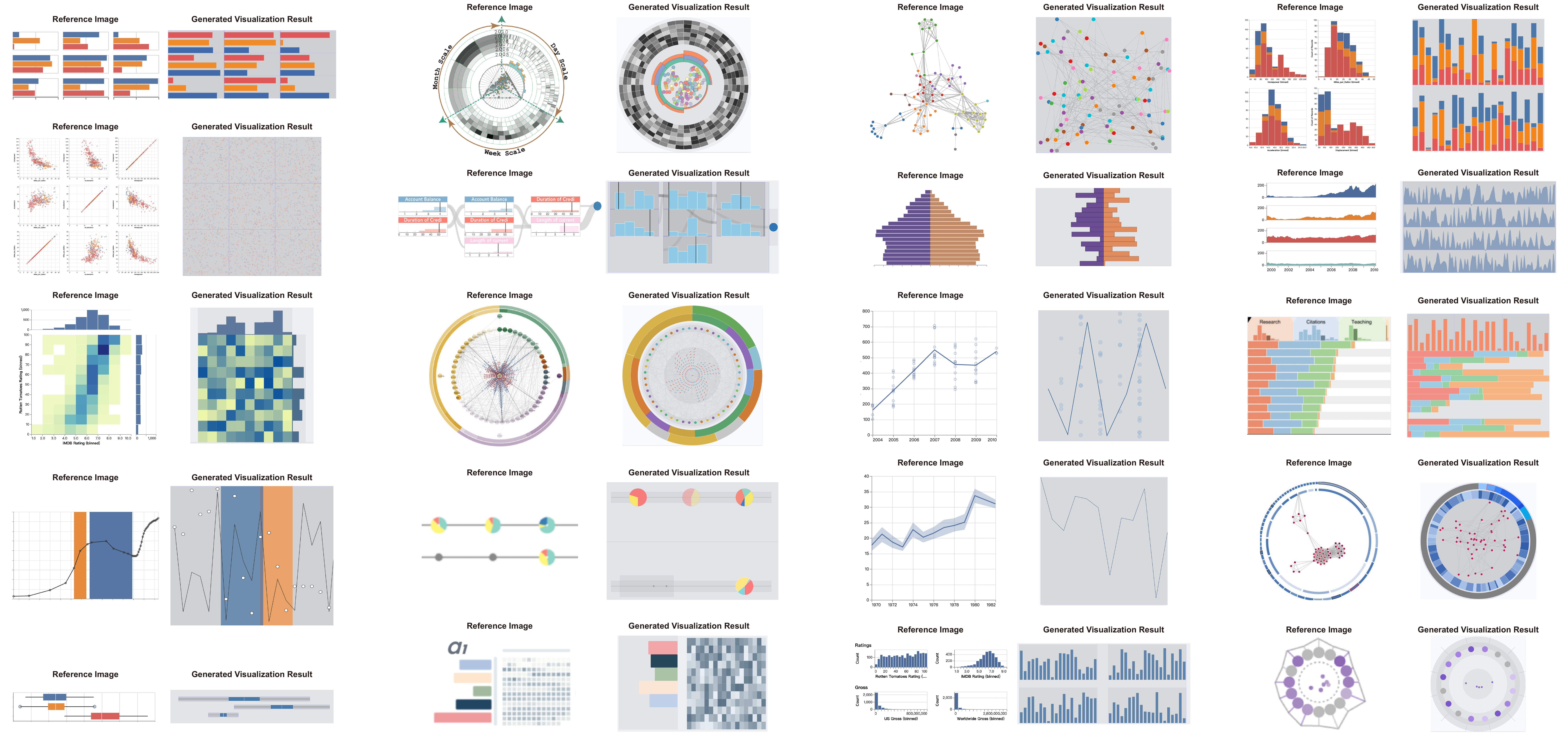}
    \caption{
    The 20 composite visualizations were used for the Quantitative Evaluation. The original input image-based visualizations are shown on the left, and the automatically reproduced results are on the right.
    }
    \label{fig: composite}
\end{figure}

Figure~\ref{fig: composite} shows the 20 composite visualizations provided to the participants in the quantitative evaluation (Section.~\ref{sec: composite_evaluation}).
The original input image-based visualizations are shown on the left, and the automatically reproduced results are on the right.

\section*{B. Domain-specific Language for Visualization Reproduction}
In the appendix, we provide a detailed introduction to the attributes in our Domain-specific Language (DSL). We use the structured model outputs~\footnote{\url{https://platform.openai.com/docs/guides/structured-outputs}} provided by OpenAI to constrain the output format. 
\begin{tcolorbox}[
  title={Hierarchical Containers},
  colback=gray!5,
  colframe=gray!60,
  fonttitle=\bfseries,
  breakable
]
\ttfamily

\textit{\textbf{container\_id}}: the unique identifier for each container, like 0-1, 0-2, 0-1-2; if the container is a template container, it should end with a letter like 0-1-a.

\textit{\textbf{description}}: an explanation of the visual components represented by this container, which is used for LLM to understand and locate the target container that needs to be parsed.

\textit{\textbf{coordinate}}: the coordinate type of the container, which could be filled by ``polar'' or ``Cartesian''.

\textit{\textbf{coordinate\_system}}: the parameters of the coordinate system, including x1, x2, y1, y2 for Cartesian coordinate and r1, r2, a1, a2, cx, cy for polar coordinates. For the template container, it shows the boundary box of all instances.

\textit{\textbf{if\_leaf}}: true or false, a mark for leaf containers with only one type of visual marks.

\textit{\textbf{components}}: sub-containers of a container, showing the hierarchical structure of our DSL.

\textit{\textbf{data\_specification}}: Only the root container includes a \textit{data\_specification}, which is an object where each key corresponds to a \textit{container\_id} and each value specifies the associated \textit{data\_specification} details. Non-leaf containers do not have \textit{data\_specification}.

\end{tcolorbox}

\begin{tcolorbox}[
  title={Data Specification of each Container},
  colback=gray!5,
  colframe=gray!60,
  fonttitle=\bfseries,
  breakable
]
\ttfamily

\textit{\textbf{mark\_specification}}: a definition of visual marks of the leaf container.

\begin{itemize}
    \item \textit{\textbf{mark\_type}}: the type of visual marks, including circle, arc, rectangle, line, band, and area. 
    \item \textit{\textbf{is\_link\_mark}}: true or false, marking if the mark is regarded as a node mark or a link mark.
    \item \textit{\textbf{link\_mark\_type}}: \textit{no\_link} indicates a node mark; \textit{group\_type} indicates that a link mark is defined by explicit control points; and \textit{node\_link} indicates that a link mark is defined by the positions of other visual marks or containers.
    \item \textit{\textbf{link\_number}}: the number of mocked links when \textit{link\_mark\_type} is \textit{node\_link}.
\end{itemize}

\textit{\textbf{data\_structure}}: a definition of the data structure of visual marks.
\begin{itemize}
    \item \textbf{\textit{data\_type}}: \textit{1D\_List}, \textit{2D\_List}, or \textit{2D\_Matrix}. 1D list represents a collection of data points without
visual grouping patterns, where each data point corresponds to an independent visual element; a 2D matrix describes data organized into groups of equal size, where each group forms a complete visual entity; a 2D list is used when the data points are grouped, but the group sizes vary across items.
    \item \textbf{\textit{data\_size}}: \textit{\textbf{primary}} and \textit{\textbf{secondary}} specify the \textit{\textbf{dimension}} (e.g., x, y, angle, and radius), placing the groups or items in the group. The \textit{primary} axis specifies how groups of data points are distributed, corresponding to the outer grouping structure. The \textit{secondary} axis specifies how individual data points within each group are arranged, determining the internal composition of each group. The \textit{\textbf{number}} attribute of primary or secondary indicates the number of elements (i.e., groups or items within groups) along each dimension, assisting data mocking and layout reasoning.
\end{itemize}

\textit{\textbf{layout\_specification}}: layout definition for visual marks and control points for each of the used axes (e.g., \textbf{\textit{x}}, \textit{\textbf{y}}, \textit{\textbf{radius}}, and \textit{\textbf{angle}}) or \textit{\textbf{source}} and \textit{\textbf{target}} nodes for \textit{node\_link} mark.
For each axis, we define the following attributes:

\begin{itemize}
    \item \textit{\textbf{stacking}}: true or false, indicating whether elements are stacked along this axis.
    \item \textit{\textbf{stacking\_direction}}: min, middle, or max, showing the direction of stacking, applicable only when \textit{stacking} is true.
    \item \textit{\textbf{subdividing}}: true or false, indicating if the stacked elements fill the entire range of the axis, applicable only when \textit{stacking} is true; if \textit{subdividing} is true, \textit{stacking\_direction} is useless.
    \item \textit{\textbf{size\_uniform}}: true or false, indicating if all elements have the same size on this axis.
    \item \textit{\textbf{size\_range}}: [min, max] from 0 to 100, showing the proportion range of size along the axis; if \textit{size\_uniform} is true, min and max are the same. 
    \item \textit{\textbf{anchor}}: min, middle, or max, indicating the used anchor for each element to decide the position, only applicable when \textit{stacking} is false.
    \item \textit{\textbf{anchor\_distribution}}: \textit{fixed\_value}, \textit{uniform\_interval}, or \textit{flexible}, indicating the layout pattern of elements along the axis.
    \item \textit{\textbf{anchor\_interval}}: Specifies the interval length between anchors, ranging from 0 to 100, and is applicable only when \textit{anchor\_distribution} is set to \textit{uniform\_interval}.

    \item \textit{\textbf{anchor\_start}}: the start position of anchors, applicable when \textit{anchor\_distribution} is \textit{fixed\_value} or \textit{uniform\_interval}.
    
\end{itemize}

\textit{\textbf{non\_layout\_specification}}: defines attributes irrelevant to position and size, such as \textit{\textbf{fill}}, \textbf{\textit{stroke}}, \textit{\textbf{stroke\_width}}, and \textit{\textbf{opacity}}. 

\begin{itemize}
    \item \textit{\textbf{scale}}: \textit{fix}, \textit{linear}, \textit{categorical}, \textit{ordinal\_primary}, or \textit{ordinal\_secondary}, the type of mapping functions.
    \item \textit{\textbf{fix}}: the fixed value of this attribute, only applicable when \textit{scale} is \textit{fix}.
    \item \textit{\textbf{linear}}: [min, max], a range for this attribute, only applicable when \textit{scale} is \textit{linear}.
    \item \textit{\textbf{options}}: a list of optional values of this attribute, applicable for \textit{categorical}, \textit{ordinal\_primary}, or \textit{ordinal\_secondary}.
\end{itemize}

\end{tcolorbox}

\section*{C. Prompts for DSL generation}
In the appendix, we provide the detailed prompts for the three steps of the DSL generator.
For Step 1: Parsing Hierarchical Containers, the input is the visualization image and a prompt as follows:

\begin{tcolorbox}[
  title={Prompt for Parsing Hierarchical Containers},
  colback=gray!5,
  colframe=gray!60,
  fonttitle=\bfseries,
  breakable
]
\ttfamily
\newcommand{\code}[1]{\texttt{#1}}

You are a domain-specific language (DSL) parsing assistant for graphics and data visualization.

Your goal is to parse the visual layout of a given image step by step and output a structured DSL description.

Please analyze the visualization by decomposing it purely based on its visual spatial layout first — focus on the spatial grouping, concentric layers, angular sectors, and element density or geometry. Do not assume any data encoding (e.g., time, value, categories) or semantic meaning until the visual structure is clearly separated and described. Prioritize visual segmentation based on shape, position, size, and graphical patterns.

Always parse and describe data-encoding marks first (circle, rectangle, arc, line, band, area).
Ignore decorative or annotation elements (labels, arrows, legends, dashed lines, axes and tick marks, explanatory text, callout boxes, margin strip). Do not include them in the output, but if the line, band, or area is used to encode data, connect two visual components, or indicate necessary information, it should be included in the output.

\textbf{(important)} Do not miss any data-encoded visual components in the visual design.

Connecting bands or lines between visual components should be considered as
an individual visual component (one leaf container for all bands/lines),
whose coordinate ranges depend on the nodes it connects.

\textbf{Step 1: Determine the Coordinate System}

\begin{itemize}
  \item First, identify the coordinate system of the entire visual design (root container):
  \begin{itemize}
    \item \textbf{Cartesian system}:
    \begin{itemize}
      \item Define the bottom-left corner \code{(x1, y1)} and the top-right corner \code{(x2, y2)} to establish the full coordinate space.
      \item default: \code{(x1: 0, y1: 0, x2: 100, y2: 100)} and \code{(x2, y2)} depends on the rate of width and height of the visual design.
    \end{itemize}
    \item \textbf{Polar system}:
    \begin{itemize}
      \item Define the polar center \code{(cx, cy)}, the angular axis range \code{(a1, a2)}, and the radial axis range \code{(r1, r2)}.
      \item default: \code{(cx: 0.5, cy: 0.5, a1: 0, a2: 360, r1: 0, r2: 1)}
    \end{itemize}
  \end{itemize}
  \item circular design uses polar coordinates, and Cartesian design uses Cartesian coordinates.
  \item only allow polar coordinates into a Cartesian container, and not allow Cartesian coordinates into a polar container.
\end{itemize}

\textbf{Step 2: Decompose into Visual Components}

Before splitting, first decide whether the design is a single basic chart or multiple sub-views.

\begin{itemize}
  \item If the whole visualization is a single basic chart composed of repeated marks of one type in one shared coordinate system (for example: a simple bar chart, radial bar chart with several concentric arcs, pie chart, donut chart, or scatter plot), you \textbf{MUST} keep the entire chart as one single leaf container:
  \begin{itemize}
    \item set \code{if\_leaf: true},
    \item set \code{mark\_type} to that mark type,
    \item and set \code{components: null}.
  \end{itemize}
  In this case, you are \textbf{not allowed} to create any child containers.
  \item Only split the design into multiple sub-containers when there are clearly different visual groups, such as:
  \begin{itemize}
    \item different mark types (e.g., bars + line + area),
    \item visually separated regions (e.g., small multiples, multiple panels),
    \item or an overlaid band/connector that should be treated as its own layer.
  \end{itemize}
  \item Check whether the visual design can be split into \textbf{non-overlapping sub-visual components} along one axis of the chosen coordinate system.
  \item If possible, divide the design into multiple \textbf{sub-containers}, each representing one visual component. If the visualization is a basic chart with only one type of visual mark, such as a line, rectangle, circle, arc, or area, do not decompose it further.
  \item Assign a distinct coordinate subspace to each container.
  \item Example for polar coordinates:
  \begin{itemize}
    \item Split along the radial \code{(r)} axis into three containers:
    \item \code{container0} covering \code{(r: 0–0.3)}
    \item \code{container1} covering \code{(r: 0.3–0.6)}
    \item \code{container2} covering \code{(r: 0.6–1.0)}
  \end{itemize}
  \item Example for Cartesian coordinates:
  \begin{itemize}
    \item Split along the x-axis into three containers:
    \item \code{container0} covering \code{(x: 0–30)}
    \item \code{container1} covering \code{(x: 30–60)}
    \item \code{container2} covering \code{(x: 60–100)}
  \end{itemize}
  \item Generate a detailed description for each container to explain the visual component it represents, including the visual elements and their relative positions.
  \item Decompose a container into multiple containers if it can be split along one axis.
  \item If it can be split along the two axes like a grid, split it into a grid and assign a distinct coordinate subspace to each container similarly.
  \item If there are overlapping containers, split them into multiple containers, and assign a distinct coordinate subspace to each container similarly.
  \item For example:
  \begin{itemize}
    \item If a visual design is an enhanced scatter plot, with each scatter plot as a new polar-based glyph, each glyph as a new container with a polar coordinate system.
  \end{itemize}
\end{itemize}

\textbf{Step 3: Recursively Decompose Sub-Containers}

\begin{enumerate}
  \item \textbf{Check if further decomposition is possible}:
  \begin{itemize}
    \item Inspect whether the current sub-container can be divided into smaller, non-overlapping sub-components based on its own coordinate system.
    \item Possible division dimensions include radial \code{(r)} bands, angular \code{(a)} sectors, or rectangular subdivisions (if Cartesian).
  \end{itemize}
  \item \textbf{If decomposition is possible}:
  \begin{itemize}
    \item Create new child containers for each identified sub-region.
    \item Define their coordinate ranges (\code{r1}, \code{r2}, \code{a1}, \code{a2} in polar; or \code{(x1, y1)–(x2, y2)} in Cartesian).
    \item Update the parent container's \code{components} to include these new sub-containers.
  \end{itemize}
  \item \textbf{If no further decomposition is possible}:
  \begin{itemize}
    \item Treat this sub-container as a leaf node.
    \item Add the \code{mark\_type} to the container.
  \end{itemize}
  \item must not overly decompose the same atomic visual marks:
  \begin{itemize}
    \item If a container only includes one type of atomic visual marks that share the same encoding and layout rule (e.g., multiple arcs in a radial bar chart, multiple bars in a grouped bar chart, multiple points in a scatter plot), treat them as \textbf{one single leaf container}. Do NOT split them into one container per mark.
  \end{itemize}
\end{enumerate}

\textbf{Repeat recursively until every container is either:}
\begin{itemize}
  \item Fully decomposed into the same atomic visual marks: circle, rectangle, arc, line, band, area.
  \item Confirmed as indivisible (leaf container), a different mark type could be further divided into multiple containers.
  \item Assign a unique \code{container\_id} to each container.
  \item Specify the coordinate system for each container with concrete values.
  \item if the container is a leaf container, set \code{if\_leaf} to true, and set the atomic visual marks, like \code{mark\_type}: circle, rectangle, arc, line, band, area; must not wrongly set arcs in polar coordinates as rectangles.
  \item If the container has components, mark it the \code{if\_leaf} as false.
  \item Please do not miss any visual components in the visual design; double-check all visual elements are included in the DSL.
  \item If the container includes only one type of visual mark can be layout with one \textbf{same} rule, mark the container as a leaf container, and no need to further decompose it. For example, do not decompose a stacked bar chart into individual bars, even if they have different fill colors.
  \item all leaf container \textbf{must} have a \code{mark\_type}. double check all leaf containers have a \code{mark\_type}.
\end{itemize}

\end{tcolorbox}

For Step 2: Extracting Template Containers, this stage uses two prompts in sequence. First, ``template\_parsing\_1'' takes the initial structure and cleans the DSL, normalizes container attributes, and detects candidate template containers, producing a TemplateParseOutput that includes cleaned\_dsl and a template\_index list. 

\begin{tcolorbox}[
  title={Prompt for Extracting Template Containers (template\_parsing\_1)},
  colback=gray!5,
  colframe=gray!60,
  fonttitle=\bfseries,
  breakable
]
\ttfamily
\newcommand{\code}[1]{\texttt{#1}}

You are a highly capable, thoughtful, and precise assistant.
Your task is to parse a DSL container tree, identify repeated composite units, and merge them into data-driven template containers.

Return ONLY JSON as final output.

Think step by step, but only output the final JSON.

\textbf{Step 1: Analyze the DSL Structure}

The following DSL describes containers and their components:
\verb|{structure_result}|

\textbf{Step 2: Identify Template Patterns}

\textbf{Template granularity (very important)}
\begin{itemize}
  \item A data-driven template container should correspond to the \textbf{smallest repeated composite unit} whose children are leaf mark-containers (e.g., one small panel containing bars and a reference line).
  \item If both a larger group (e.g., a whole column of panels) and a smaller unit (one panel) are repeated:
  \begin{itemize}
    \item \textbf{Always choose the smaller composite (the panel) as the template container.}
    \item The larger grouping (columns / multiple rows) must be represented only in data patterns (later) and NOT as another template container.
  \end{itemize}
  \item Within one spatial region, do NOT create nested template containers that are generated by the same higher-level data pattern (no ``template of templates'').
  \item When deciding the root for a data-driven template:
  \begin{itemize}
    \item Prefer the \textbf{lowest} container whose children are leaf mark-containers (e.g., one small panel).
    \item Do NOT add an extra template container that only groups multiple template instances.
    \item Replace the description of the template container with a short summary of the repeated visual elements, its outer bounding box, and its layout pattern.
  \end{itemize}
\end{itemize}

\textbf{Merging rules}
\begin{itemize}
  \item Follow a top-down approach, traversing top containers first and then inner containers.
  \item Identify repeated structures.
  \item \textbf{Remove} repeated structures, keeping one data-driven template container that stands for all repeated instances.
  \item Replace the \code{container\_id} with a variable where it is repeated (e.g., \code{0-a}, \code{0-b}, \code{0-c}):
  \begin{itemize}
    \item Use different variables \code{a}, \code{b}, \code{c}, etc. for different template patterns.
    \item Do not use the same variable for different template patterns.
  \end{itemize}
  \item For each data-driven template container:
  \begin{itemize}
    \item Its \code{coordinate\_system} should be the \textbf{bounding box of all its instances} (the full region where all repeated units appear).
    \item However, \textbf{one instance of this template still corresponds to a single repeated visual unit} (e.g., one small histogram panel).
    \item Inside this template container, you should only keep the structure of \textbf{one representative instance} (one panel), not all of them.
  \end{itemize}
  \item Only merge containers when:
  \begin{itemize}
    \item They have the same visual elements and the same data encodings.
    \item Their only difference is spatial position and underlying data, not layout logic.
    \item They can be generated from the same data pattern (i.e., one unified data-driven template).
  \end{itemize}
  \item (Important) Do not merge containers when:
  \begin{itemize}
    \item Visual elements differ,
    \item Data encodings differ,
    \item Semantics differ, or
    \item Their layouts follow different rules (e.g., stacking vs. gridding, or stacking direction differs), even if the marks look similar.
  \end{itemize}
  \item Only merge existing containers when necessary. Do not introduce new hierarchy levels beyond what is needed to represent the template.
  \item Do not change \code{mark\_type} values.
  \item When you have already decided that a container is a data-driven template:
  \begin{itemize}
    \item Only analyze \textbf{one instance} of its contents to clean redundancy.
    \item Do NOT create another data-driven template inside it for repeated containers that belong to the same repetition pattern.
    \item Inside a template container, repeated leaf marks (e.g., multiple bars inside one panel) should remain as normal mark containers, NOT as new template containers.
  \end{itemize}
\end{itemize}

\textbf{Examples}

\textit{1D merge example input}
\begin{verbatim}
{{
  "container_id": "0",
  "coordinate": "cartesian",
  "coordinate_system": {{
    "x1": 0, "y1": 0, "x2": 100, "y2": 100
  }},
  "components": [
    {{
      "container_id": "0-0",
      "description": "same visual component.",
      "coordinate": "cartesian",
      "coordinate_system": {{
        "x1": 0, "y1": 0, "x2": 30, "y2": 100
      }}
    }},
    {{
      "container_id": "0-1",
      "description": "same visual component.",
      "coordinate": "cartesian",
      "coordinate_system": {{
        "x1": 30, "y1": 0, "x2": 60, "y2": 100
      }}
    }},
    {{
      "container_id": "0-2",
      "description": "different visual component"
    }}      
  ]
}}
\end{verbatim}

\textit{1D merge example output}
\begin{verbatim}
{{
  "container_id": "0",
  "coordinate": "cartesian",
  "coordinate_system": {{
    "x1": 0, "y1": 0, "x2": 100, "y2": 100
  }},
  "components": [
    {{
      "container_id": "0-a",
      "description": "template for a single repeated component, with its 
      coordinate representing the outer boundary of all repeated instances, 
      and its layout structure is a 1D_LIST.",
      "coordinate": "cartesian",
      "coordinate_system": {{
        "x1": 0, "y1": 0, "x2": 60, "y2": 100
      }}
    }},
    {{
      "container_id": "0-2",
      "description": "different visual component"
    }}
  ]
}}
\end{verbatim}

\textit{2D merge example input}
\begin{verbatim}
{{
  "container_id": "0",
  "coordinate": "cartesian",
  "description": "example root",
  "coordinate_system": {{
    "x1": 0, "y1": 0, "x2": 100, "y2": 100
  }},
  "components": [
    {{
      "container_id": "0-0",
      "description": "the first column has three small panels",
      "coordinate": "cartesian",
      "coordinate_system": {{
        "x1": 0, "y1": 0, "x2": 30, "y2": 90
      }}            
      "components": [
        {{
          "container_id": "0-0-0",
          "description": "the first panel in the first column",
          "coordinate_system": {{
            "x1": 0, "y1": 0, "x2": 30, "y2": 30
          }}
        }},
        {{
          "container_id": "0-0-1",
          "description": "the second panel in the first column",
          "coordinate_system": {{
            "x1": 0, "y1": 30, "x2": 30, "y2": 60
          }}    
        }},
        {{
          "container_id": "0-0-2",
          "description": "the third panel in the first column",
          "coordinate_system": {{
            "x1": 0, "y1": 60, "x2": 30, "y2": 90
          }}
        }}
      ]
    }},
    {{
      "container_id": "0-1",
      "description": "the second column has two small panels, the panels 
      are the same as the first column",
      "coordinate": "cartesian",
      "coordinate_system": {{
        "x1": 30, "y1": 0, "x2": 60, "y2": 60
      }},
      "components": [
        {{
          "container_id": "0-1-0",
          "description": "the first panel in the second column",
          "coordinate": "cartesian",
          "coordinate_system": {{
            "x1": 30, "y1": 0, "x2": 60, "y2": 30
          }}
        }},
        {{
          "container_id": "0-1-1",
          "description": "the second panel in the second column",
          "coordinate": "cartesian",
          "coordinate_system": {{
            "x1": 30, "y1": 30, "x2": 60, "y2": 60
          }}
        }}   
      ]
    }}
  ]
}}
\end{verbatim}

\textit{2D merge example output}
\begin{verbatim}
{{  
  "container_id": "0",
  "coordinate": "cartesian",
  "description": "example root",
  "coordinate_system": {{
    "x1": 0, "y1": 0, "x2": 100, "y2": 100
  }},
  "components": [
    {{
      "container_id": "0-a",
      "description": "template for a single repeated panel, laid out 
      in a 2D grid; its coordinate represents the outer boundary of all 
      repeated panel instances.",
      "coordinate": "cartesian",
      "coordinate_system": {{
        "x1": 0, "y1": 0, "x2": 60, "y2": 90
      }},
      "components": [...]
    }}  
  ]
}}
\end{verbatim}

\textbf{Step 3: Output requirements}

You must output a JSON object with TWO fields:

\code{cleaned\_dsl}: the merged DSL with template containers (no \code{template\_data\_specifica-
tion} here).

\code{template\_index}: an array that records which original containers were merged into each template.

\textbf{Schema:}
\begin{verbatim}
{{
  "cleaned_dsl": {{ ... }},
  "template_index": [
    {{
      "template_id": "0-a",
      "instance_ids": ["0-0", "0-1"],
      "instance_bboxes": [
        {{
          "x1": 0, "y1": 0, "x2": 30, "y2": 100
        }},
        {{
          "x1": 30, "y1": 0, "x2": 60, "y2": 100
        }}
      ]
    }}
  ]
}}
\end{verbatim}

\code{template\_id} is the \code{container\_id} of the merged template container (e.g., ``0-a'').

\code{instance\_ids} are the original \code{container\_ids} merged into this template.

\code{instance\_bboxes} are the original \code{coordinate\_system} values for each instance.

Only include templates that correspond to repeated containers.

Return ONLY this JSON. No extra text.
\end{tcolorbox}

Second, for each entry in template\_index, we apply ``template\_parsing\_2'' to extract a TemplateDataSpecification describing the data schema and binding required by the template container. The resulting specifications are consolidated under template\_data\_specification within the cleaned DSL.

\begin{tcolorbox}[
  title={Prompt for Extracting Template Containers (template\_parsing\_2)},
  colback=gray!5,
  colframe=gray!60,
  fonttitle=\bfseries,
  breakable
]
\ttfamily
\newcommand{\code}[1]{\texttt{#1}}
You are a highly capable, thoughtful, and precise assistant.
Your task is to infer the \code{template\_data\_specification} for a data-driven template container, based on a cleaned DSL and a \code{template\_index} describing which original containers were merged.

Return ONLY JSON as final output.

Think step by step, but only output the final JSON.

\textbf{Inputs}

- Original DSL:
\verb|{structure_result}|

- Cleaned DSL with template containers:
\verb|{cleaned_dsl}|

- Template index mapping templates to their original instances:
\verb|{template_index}|

\code{template\_index} has:
\begin{itemize}
  \item \code{template\_id}: the \code{container\_id} of the template container in the cleaned DSL (e.g., ``0-a'').
  \item \code{instance\_ids}: original \code{container\_ids} that were merged into this template.
  \item \code{instance\_bboxes}: their original \code{coordinate\_system} bounding boxes.
\end{itemize}

Each template container corresponds to a repeated composite unit, and each instance of that composite corresponds to one data item (or one cell in a 1D/2D structure).

\textbf{Step 1: Infer Data Structure}
(\code{data\_type} and \code{data\_size})

For the \code{template\_index}:

1. Treat each instance in \code{instance\_bboxes} as one data instance.

2. \textbf{Select Data Type}
\begin{itemize}
  \item \code{1D\_LIST} $\rightarrow$ containers arranged equally along a single dimension (horizontal row, vertical column, or circular ring).
  \item \code{2D\_MATRIX} $\rightarrow$ instances form a regular grid (uniform number of rows $\times$ columns).
  \item \code{2D\_LIST} $\rightarrow$ groups with varying item counts per group.
  \item Always choose the \textbf{simplest} structure that fits all instances.
\end{itemize}

3. \textbf{Determine Dimensions}
\begin{itemize}
  \item Use positions in bounding boxes to decide which dimension(s) drive repetition:
  \begin{itemize}
    \item primary dimension = outer repetition axis.
    \item secondary dimension = inner repetition within each group (for 2D structures).
  \end{itemize}
  \item If only one axis is used, it is the primary.
\end{itemize}

4. \textbf{Count Elements}
\begin{itemize}
  \item \code{primary.number} = the number of groups or instances along the primary dimension.
  \item \code{secondary.number}:
  \begin{itemize}
    \item integer for \code{2D\_MATRIX} (consistent count per group).
    \item integer array for \code{2D\_LIST}, listing group-wise counts.
  \end{itemize}
  \item Add \code{explanation} ($\le$ 20 words) for both primary and secondary (if present), briefly justifying the numbers.
\end{itemize}

\textbf{Step 2: Infer Layout Specification}
(\code{layout\_specification})

For each template, use the spatial distribution of its \code{instance\_bboxes}.

For each dimension that influences layout (\code{x}, \code{y}, \code{radius}, \code{angle}), infer:
\begin{itemize}
  \item \code{stacking}: whether elements accumulate along this dimension.
  \item \code{stacking\_direction}: \code{'min'} \textbar{} \code{'max'} \textbar{} \code{'middle'}.
  \begin{itemize}
    \item For \code{x}: \code{min} is left, \code{max} is right.
    \item For \code{y}: \code{min} is bottom, \code{max} is top.
  \end{itemize}
  \item \code{anchor}: \code{'min'} \textbar{} \code{'max'} \textbar{} \code{'middle'} \textbar{} \code{'stacking\_decided'}.
  \item \code{subdividing}: \code{true} only if stacking and the instances essentially fill the whole dimension.
  \item \code{2d\_flatten}: whether hierarchical groups are flattened along this dimension.
  \item \code{size\_uniform}: whether instance sizes are consistent in this dimension.
  \item \code{size\_range}: \code{[min, max]} in \textbf{0--100} relative units (normalize using the template’s overall bounding box from the cleaned DSL).
  \item \code{anchor\_distribute}: \code{'fixed\_value'} \textbar{} \code{'uniform\_interval'} \textbar{} \code{'flexible'}.
  \item \code{anchor\_start}: first anchor position (0--100).
  \item \code{anchor\_interval}: spacing between anchors (0--100).
  \item \code{number}: element count or array, aligning with the inferred data structure.
\end{itemize}

\textbf{Normalization Rules}
\begin{itemize}
  \item Normalize positions (e.g., \code{anchor\_start}, \code{anchor\_interval}) to \textbf{[0--100]} along each dimension.
  \begin{itemize}
    \item If \code{anchor\_distribute = 'fixed\_value'}, \code{anchor\_start} is the exact position.
    \item If \code{anchor\_distribute = 'uniform\_interval'}, \code{anchor\_start} is the first position and \code{anchor\_interval} is the spacing.
    \item Ensure \code{anchor\_start + number * anchor\_interval <= 100}.
  \end{itemize}
  \item Normalize size values (\code{size\_range}) to \textbf{[0--100]} relative to the template’s bounding box.
\end{itemize}

\textbf{Ambiguity Handling}
When evidence is uncertain:
\begin{itemize}
  \item Prefer simpler \code{data\_type} (\code{1D\_LIST} $<$ \code{2D\_MATRIX} $<$ \code{2D\_LIST}).
  \item Treat spacing as \code{'uniform\_interval'} when variance $\le$ 5\%.
  \item If both \code{x} and \code{y} could be primary, pick the one with larger inter-group spacing ratio.
  \item Omit unused dimensions from \code{layout\_specification}.
\end{itemize}

\textbf{Step 3: Output Requirements}

You must output a single JSON object:

\begin{verbatim}
{
  "container_id": "0-a",
  "data_structure": {
    "data_type": "1D_LIST | 2D_MATRIX | 2D_LIST",
    "data_size": {
      "primary": {
        "number": 0,
        "dimension": "x | y | radius | angle | [dimension]",
        "explanation": "Explain the reason for choosing this number"
      },
      "secondary": {
        "number": 0,
        "dimension": "x | y | radius | angle",
        "explanation": "Explain the reason for choosing this number"
      }
    }
  },
  "layout_specification": {
    "x": {
      "stacking": false,
      "stacking_direction": "min | max | middle",
      "anchor": "min | max | middle | stacking_decided",
      "subdividing": false,
      "2d_flatten": false,
      "size_uniform": false,
      "size_range": [0, 0],
      "anchor_distribute": "fixed_value | uniform_interval | flexible",
      "anchor_interval": 0,
      "anchor_start": 0
    },
    "y": {
      "stacking": false,
      "stacking_direction": "min | max | middle",
      "anchor": "min | max | middle | stacking_decided",
      "subdividing": false,
      "2d_flatten": false,
      "size_uniform": false,
      "size_range": [0, 0],
      "anchor_distribute": "fixed_value | uniform_interval | flexible",
      "anchor_interval": 0,
      "anchor_start": 0
    },
    "radius": { ... },
    "angle": { ... }
  }
}
\end{verbatim}

Only include dimensions actually used in the layout.

Use exact field names and enum strings.

Return ONLY this JSON. No extra text.

\end{tcolorbox}

For Step 3: Parsing Encodings of Visual Mark, given the image and the cleaned DSL context, the prompt “mark\_parsing” targets leaf containers and parses the visual encodings (marks, channels, fields, scales). The output is a DataSpecification per leaf container. These per-container specifications are merged into the final DSL, alongside the template data specifications, to form the complete result.

\begin{tcolorbox}[
  title={Prompt for Parsing Encodings of Visual Mark},
  colback=gray!5,
  colframe=gray!60,
  fonttitle=\bfseries,
  breakable
]
\ttfamily
\newcommand{\code}[1]{\texttt{#1}}

You are a data generation assistant.\\
Your task: \textbf{Given the DSL (container tree + visual marks) and the design image, produce the JSON data specification for the target container and mark type.}\\
Follow the schema constraints strictly.

\textbf{1. Context}

\begin{itemize}
  \item DSL: \verb|{dsl}|
  \item Target:
  \begin{itemize}
    \item \code{mark\_type} = \verb|{mark_type}|
    \item \code{container\_id} = \verb|{container_id}|
  \end{itemize}
\end{itemize}

Only consider marks that belong to \verb|{container_id}|.\\
If the DSL uses a template container (with letter suffix like ``\code{\_a}''), parse \textbf{one representative instance} using its own relative coordinate.

\textbf{Output must be JSON only. No text.}

\textbf{2. Output Schema (Top-Level)}

The final returned JSON must match:
\begin{verbatim}
{
  "data_structure": {...},
  "mark_specification": {...},
  "layout_specification": {...},
  "non_layout_specification": {...}
}
\end{verbatim}

\textbf{3. Step 1 --- Determine Data Structure}

\textbf{3.1 Choose \code{data\_type}}\\
Three valid categories:

\textbf{(1) \code{1D\_LIST}}\\
A single repeated sequence. dimension refers to the layout dimensions involved.\\
Examples:
\begin{itemize}
  \item A simple bar chart along x $\rightarrow$ dimension = ['x']
  \item A scatter plot $\rightarrow$ dimension = ['x','y'] // circles distribute data points in two dimensions.
\end{itemize}

Format:
\begin{verbatim}
{
  "primary": {
    "number": <int>,
    "dimension": "x" | "y" | ["x","y"] | "radius" | "angle" | ["radius","angle"]
  }
}
\end{verbatim}

\textbf{(2) \code{2D\_MATRIX}}\\
A regular grid: \textbf{rows $\times$ columns} where both counts are consistent.\\
primary dimension means the dimension that is used to layout the group, and secondary dimension means the dimension that is used to layout the item in each group.\\
Grouped bar chart $\rightarrow$ primary dimension = 'x', secondary dimension = 'x', since bar group subdivide x dimension and each bar in a group subdivide x dimension.

Format:
\begin{verbatim}
{
  "primary": {
    "number": <int>,
    "dimension": "x" | "y" | "radius" | "angle"
  },
  "secondary": {
    "number": <int>,
    "dimension": "x" | "y" | "radius" | "angle"
  }
}
\end{verbatim}

\textbf{(3) \code{2D\_LIST}}\\
A grid where \textbf{each row/column has a different number of items}.\\
\code{secondary.number} must be an ascending array.\\
The array length equals the primary.number, and each value represents the number of elements in each group.

Format:
\begin{verbatim}
{
  "primary": {
    "number": <int>,
    "dimension": "x" | "y" | "radius" | "angle"
  },
  "secondary": {
    "number": [int, int, ...],
    "dimension": "x" | "y" | "radius" | "angle"
  }
}
\end{verbatim}

\textbf{3.2 Handling link marks (very important)}

If the target mark is a link-type (line, band, area), determine:

\textbf{Case A. group\_type (continuous polyline/curve)}\\
Used for:
\begin{itemize}
  \item line charts, area charts, radar, band curves...
\end{itemize}

Properties:
\begin{itemize}
  \item Require 2D\_MATRIX or 2D\_LIST.
  \item Each group = one link. Draw a link-mark through a set of control points.
  \item the data structure remains the same, but the number of groups equals the number of link marks. Use a group of data points for one link mark and layout the control points for each link mark.
  \item group\_link\_direction: 'x' | 'y' | 'radius' | 'angle',
  \begin{itemize}
    \item 'x' $\rightarrow$ link direction is along x dimension
    \item 'y' $\rightarrow$ link direction is along y dimension
    \item 'radius' $\rightarrow$ link direction is along radius dimension
    \item 'angle' $\rightarrow$ link direction is along angle dimension
  \end{itemize}
  \item is\_width\_encoded\_data: True/False
  \begin{itemize}
    \item e.g., area chart = True
    \item line chart = False
  \end{itemize}
\end{itemize}

\textbf{Case B. node\_link\_type (graph edges)}\\
Properties:
\begin{itemize}
  \item link\_mark\_type = 'node\_link\_type'
  \item link\_number: number of link marks
  \item source: container ids supplying source nodes
  \item target: container ids supplying target nodes
\end{itemize}

Including only one leaf container or template container is allowed, indicating the link marks linking the node marks or intances within this container.\\

\textbf{Case C. no\_link}\\
If not a link mark:
\begin{itemize}
  \item link\_mark\_type = 'no\_link'
  \item group\_link\_direction = null
  \item is\_width\_encoded\_data = false
\end{itemize}

\textbf{4. Step 2 --- Determine Layout Specification}

For each relevant dimension ('x', 'y', 'radius', 'angle'), analyze:

\textbf{stacking (boolean)}\\
Whether items accumulate along that dimension.

\textbf{stacking\_direction ('min' | 'max' | 'middle')}
\begin{itemize}
  \item 'min' $\rightarrow$ align to left/bottom
  \item 'max' $\rightarrow$ align to right/top
  \item 'middle' $\rightarrow$ centered
\end{itemize}

\textbf{anchor ('min' | 'max' | 'middle' | 'stacking\_decided')}\\
If stacking exists $\rightarrow$ anchor = 'stacking\_decided'.\\
Rules:
\begin{itemize}
  \item If stacking $\rightarrow$ use 'stacking\_decided'
  \item If centered $\rightarrow$ 'middle'
  \item If aligned to low end $\rightarrow$ 'min'
  \item If aligned to high end $\rightarrow$ 'max'
\end{itemize}

\textbf{subdividing (boolean)}\\
True only if stacking fills the entire dimension.

\textbf{2d\_flatten (boolean)}\\
True if groups are flattened (e.g., grouped bar charts).

\textbf{size\_range ([min, max])}\\
Relative to 0--100 units of the dimension.\\
If all sizes equal $\rightarrow$ min = max.

\textbf{size\_uniform (boolean)}\\
True if item sizes along this dimension are uniform.\\
If true, size\_range = [min, max], min = max.

\textbf{anchor\_distribute ('fixed\_value' | 'uniform\_interval' | 'flexible')}
\begin{itemize}
  \item fixed\_value $\rightarrow$ anchor\_start = constant, anchor\_interval = 0, all items have the same anchor position.
  \item uniform\_interval $\rightarrow$ specify anchor\_start + anchor\_interval, all items have the same interval between anchors.
  \item flexible $\rightarrow$ both values null
\end{itemize}
if fixed\_value, anchor\_start in 0--100 units of the dimension.\\
if uniform\_interval, anchor\_start + anchor\_interval * number of items in 0--100 units of the dimension.

\textbf{source / target}\\
Used \textbf{only} for node-link marks.\\
A list of container ids that supply source nodes or target nodes.

\textbf{5. Step 3 --- Determine non\_layout\_specification}

Allowed fields:
\begin{itemize}
  \item fill
  \item stroke
  \item opacity
  \item stroke\_width
  \item line\_type (only for: line, band, area)
  \item rx, ry (rounded corners, rectangles only)
\end{itemize}

Each field uses one of the following scales:
\begin{itemize}
  \item 'fix': fixed value
  \item 'linear': linear scale
  \item 'ordinal\_primary': ordinal scale with primary domain
  \item 'ordinal\_secondary': ordinal scale with secondary domain
  \item 'categorical': categorical scale
\end{itemize}

\textbf{5.1 Color fields (fill, stroke):}
\begin{verbatim}
{
  "scale": "...",
  "fix"?: "#RRGGBB",
  "linear"?: ["#RRGGBB", "#RRGGBB"],
  "options"?: ["#RRGGBB", ...]
}.
\end{verbatim}

\textbf{5.2 Numeric fields (stroke\_width, opacity):}
\begin{verbatim}
{
  "scale": "fix" | "linear" | "ordinal_primary" | "ordinal_secondary" | "categoric-
  al",
  "fix"?: number,
  "linear"?: [number, number],
  "options"?: [number, ...]
}
\end{verbatim}

Rules:
\begin{itemize}
  \item opacity $\in$ [0,1]
  \item stroke\_width $\ge 0$
\end{itemize}

\textbf{5.3 line\_type}\\
Only for:
\begin{itemize}
  \item line
  \item band
  \item area
\end{itemize}

Values: "curve" | "straight"

\textbf{5.4 Constraints}
\begin{itemize}
  \item Choose \textbf{exactly one} type of scale.
  \item Only include keys belonging to the chosen scale.
  \item Do not emit keys for unused fields.
  \item Colors must be \#RRGGBB.
\end{itemize}

\textbf{6. FINAL OUTPUT FORMAT}

Return \textbf{only JSON} following:
\begin{verbatim}
{
  data_structure: {
    data_type: '1D_LIST' | '2D_MATRIX' | '2D_LIST',
    data_size: {
      primary: {
        number: number,
        dimension: 'x' | 'y' | 'radius' | 'angle' | [dimension],
        explanation: "Explain the reason briefly"
      },
      secondary?: {
        number: number | number[],
        dimension: 'x' | 'y' | 'radius' | 'angle'
      }
    }
  },

  mark_specification: {
    mark_type: 'rectangle' | 'line' | 'circle' | 'arc' | 'band' | 'area' | 'text',
    is_link_mark: boolean,
    link_mark_type: 'group_type' | 'node_link_type' | 'no_link',
    group_link_direction?: 'x' | 'y' | 'radius' | 'angle',
    link_number?: number,
    node_use_once: boolean,
    is_width_encoded_data: boolean,
    is_fully_connected: boolean,
    is_bipartite: boolean
  },

  layout_specification: {
    [dimension in 'x', 'y', 'radius', 'angle']: {
      stacking: boolean,
      stacking_direction: 'min' | 'max' | 'middle',
      anchor: 'min' | 'max' | 'middle' | 'stacking_decided',
      subdividing: boolean,
      2d_flatten: boolean,
      size_uniform: boolean,
      size_range: [number, number],
      anchor_distribute: 'fixed_value' | 'uniform_interval' | 'flexible',
      anchor_interval: number | null,
      anchor_start: number | null
    },
    source?: string[],
    target?: string[]
  },

  non_layout_specification: {
    line_type?: 'curve' | 'straight',
    stroke_width?: NumericEncoding,
    opacity?: NumericEncoding,
    fill?: ColorEncoding,
    stroke?: ColorEncoding,
    rx?: NumericEncoding,
    ry?: NumericEncoding
  }
}
\end{verbatim}
\end{tcolorbox}

\end{document}